\documentclass[twocolumn,nofootinbib,notitlepage,superscriptaddress,10pt,aps,prd]{revtex4-1}

\usepackage{graphicx,epsfig,amssymb,times} 
\usepackage{amsmath,amsfonts}
\usepackage{bm}
\usepackage{epstopdf}
\usepackage[linktocpage,colorlinks]{hyperref}
\usepackage[caption=false]{subfig}
\usepackage[usenames]{color}     
\usepackage{tikz}
\usetikzlibrary{decorations.pathmorphing}
\usepackage{float}
\usepackage{ulem}

\newcommand{\mn}{\ensuremath{{_{\mu\nu}}}}

\newcommand{\pa}[1]{\ensuremath{\partial_{#1}}}

\newcommand{\al}{\ensuremath{\alpha}}
\newcommand{\be}{\ensuremath{\beta}}
\newcommand{\ga}{\ensuremath{\gamma}}
\newcommand{\Ga}{\ensuremath{\Gamma}}

\newcommand{\na}{\ensuremath{\nabla}}

\newcommand{\bpsi}{\ensuremath{\bar{\psi}}}

\newcommand{\dvg}{\ensuremath{\dif V_{g}}}

\newcommand{\viu}[2]{\ensuremath{{e^{#1}}_{#2}}}
\newcommand{\vid}[2]{\ensuremath{{e_{#1}}^{#2}}}

\newcommand{\lag}{\mathcal{L}}

\newcommand{\lr}[1]{\left(#1\right)}
\newcommand{\lrsq}[1]{\left[#1\right]}

\newcommand{\dm}{\ensuremath{\underline{\gamma}}}

\newcommand{\Boxg}{\ensuremath{\Box_g}}

\newcommand{\dif}{\ensuremath{\textbf{\text{d}}}}
\newcommand{\bdel}{\ensuremath{\boldsymbol{\delta}}}

\newcommand{\bee}{\begin{equation}}
\newcommand{\ee}{\end{equation}}
\newcommand{\bea}{\begin{eqnarray}}
\newcommand{\eea}{\end{eqnarray}}
\newcommand{\ba}{\begin{align}}
\newcommand{\ea}{\end{align}}

\def\bs{\begin{subequations}}
\def\es{\end{subequations}}
\renewcommand{\leq}{\leqslant}

\def\cB{\mathcal{B}}
\def\cC{\mathcal{C}}

\def\bg{\textbf{g}}

\newcommand{\tia}[1]{}

\newcounter{listcounter}

\begin{document}

\title{\large Minimal coupling in presence of non-metricity and torsion}

\author{Adri\`a  Delhom}
\email{adria.delhom@uv.es}
\affiliation{Departamento de F\'isica Te\'orica and IFIC, 
Centro Mixto Universidad de Valencia - CSIC. 
Universidad de Valencia, Burjassot-46100, 
Valencia, Spain}

\begin{abstract}
We deal with the question of what it means to define a minimal coupling prescription in presence of torsion and/or non-metricity, carefully explaining while the naive substitution $\partial\to\na$ introduces extra couplings between the matter fields and the connection that can be regarded as non-minimal in presence of torsion and/or non-metricity. We will also investigate whether minimal coupling prescriptions at the level of the action (MCPL) or at the level of field equations (MCPF) lead to different dynamics. To that end, we will first write the Euler-Lagrange equations for matter fields in terms of the covariant derivatives of a general non-Riemannian space, and derivate the form of the associated Noether currents and charges. Then we will see that if the minimal coupling prescriptions is applied as we discuss, for spin 0 and 1 fields the results of MCPL and MCPF are equivalent, while for spin 1/2 fields there is a difference if one applies the MCPF or the MCPL, since the former leads to charge violation.
\end{abstract}

\maketitle

\section{Introduction}
\label{intro}
General Relativity (GR) had an important impact in the development of non-Euclidean geometries at the beginning of the twentieth century \cite{EisenhartRiem,EisenhartNRiem}. Einstein followed the (by then recent) developments of Riemann to shape his theory of gravity, and rapidly the idea that spacetime must have a Riemannian structure was naturally accepted despite the lack of direct empirical evidence \cite{Will_Book}. In recent years, a renewed interest in determining whether the space-time geometry is actually Riemannian or otherwise has emerged boosted, in part,  by the exploration of new gravity theories in the metric-affine formalism. The first step forward in this direction was already given by Cartan a few years after the birth of GR. He suggested that the torsion tensor could be introduced in the description of gravity, and he introduced the idea that it should be related to the intrisic angular-momentum of matter \cite{Cartan22,Cartan23,Cartan24,Cartan25}, but this idea was quickly forgotten given that the spin of the electrons was not already discovered.  Weyl, Einstein, and others used Cartan's idea and introduced torsion in an unsuccesful attemnpt of unifying gravity with electromagnetism \cite{weyl1,Einstein,Schrodinger,Tonnelat55,Tonnelat65}. In the mid 50's of the past century, when trying to describe the continuum limit of the micro-structure of solids, the torsion tensor was found to be related to the density of dislocation deffects present in the micro-structure \cite{Kondo1,Kondo2,Bilby1,Bilby2,Kroner1,Kroner2}. At the same time Kibble and Sciama \cite{Kibble,Sciama1,Sciama2} considered the description of gravity as a gauge theory of the Poincar\'e group, formulating the so called Einstein-Cartan-Sciama-Kibble (ECSK) theory. Soon after the formulation of ECSK theory as a gauge theory, Hehl and collaborators \cite{Hehl1,Hehl2,Hehl3,Hehl4,Hehl5,Hehl6} worked out its geometrical formulation\footnote{Also see \cite{Hehl:1994ue} for a generalization to a gauge theory of the affine group.}, showing how the spin of particles was related to torsion in ECSK, being torsion sourced by the spin-density \cite{Hehl6,Hehl7,Hehl8}. The observable consequences of torsion have been considered in several works, concluding that they are generally suppressed by a high energy scale and therefore practically unobservable at low energies, unless a scenario with high density of spin is considered \cite{Hehl7,Hehl8,Saphiro1,Saphiro2,Saphiro3}. However, it has been pointed out that torsion could have important role in early universe cosmology, as it could prevent the Big Bang singularity \cite{Poplawski}. Torsion-based theories have also received much interest recently in the context of teleparallel gravity and its functional extensions \cite{Saridakis_Capozziello_de_Laurentis}. Up to date, the different attempts looking for the experimental detection of torsion have given negative results \cite{will}. \\

Whereas the observable effects of torsion have been fairly well studied in gravitational contexts, the other signature of non-Riemannian geometry\footnote{Here we use Riemannian to imply that the connection is metric-compatible, and non-Riemannian for otherwise. Notice that Mathematicians have another meaning for Riemannian, which is related to the signature of the metric. Our metrics will be assumed to have Lorentzian signature.}, which is non-metricity, has not been analyzed as deeply (see \cite{ReviewCantata} for a short review of these results). However, there has recently been renewed interest analysing several theories that feature non-metricity and/or torsion, such as teleparallel and symmetric teleparallel theories \cite{Hayashi:1979qx,Nester:1998mp,Obukhov:2002tm,Poltorak:2004tz,Adak:2006rx,Aldrovandi:2013wha,Maluf:2013gaa,Mol:2014ooa,Krssak:2015oua,BeltranJimenez:2017tkd,Combi:2017crv,BeltranJimenez:2018vdo,Koivisto:2018loq,Lobo:2019xwp,Jimenez:2019tkx,Jimenez:2019ghw,Krssak:2018ywd,BeltranJimenez:2019tjy}, Ricci-based gravity thoeries (which encompass $f(R))$ or Born-Infeld gravity) \cite{Capozziello:2007tj,Olmo:2011uz,ReviewBI,Afonso:2018bpv,Afonso:2018hyj,Afonso:2018mxn,Letter,Delhom:2019wir,Delhom:2019zrb,BeltranJimenez:2019acz, GhostsLargo}, or others \cite{Capozziello:2015lza,Galtsov:2018xuc,delaCruz-Dombriz:2018aal,Harko:2018gxr,Xu:2019sbp}. Some of these works put forward that gravity theories with non-metricity can avoid the spacetime singularities present in GR already at a classical level \cite{Palageod,BGKM}. It has also been suggested that non-Riemannian geometries could be successful in being a low-energy effective description of theories of quantum gravity, as they could be more suited in accounting for features of a quantum-spacetime that may exist at high energies \cite{CrystalClear,SinghOlmo,Hoss,Letter,Queiruga:2019ike}. Hence, that the existence of non-metricity should be experimentally probed at different energy scales. In order to probe non-metricity, we must first understand its experimental consequences. Recent works on metric-affine theories of gravity show how non-metricity could have important effects in scenarios with very high energy-density, giving rise for instance to effective particle interactions or shifts in the energy levels of atomic systems \cite{Letter,Delhom:2019wir}, which can be used to constrain the energy scale at which Ricci-based gravity theories depart form GR. Nonetheless, it remains an open question whether Riemannianity (i.e. absence of torsion and non-metricity) extends to higher energy regimes or if it is a low energy limit of a more general non-Riemanian spacetime structure. \\

 Other current lines of research also investigate the possible implications of spacetime non-metricity in classical trajectories or in the definition of geometric clocks assuming that test bodies follow affine geodesics \cite{Avalos:2016unj,Lobo:2018zrz,Delhom:2020vpe} . Nonetheless, this assumption should be derived from the geometrical optics approximation of the field equations that describe matter at a fundamental level, and it is not yet clear how to do so (see \cite{GhostsLargo} for a discussion). In order to derive this limit, one must first understand the different ways in which matter fields couple to generic torsion and non-metricity tensors. Some possible ways of coupling geometry and matter in Riemann-Cartan space-times (includding minimal coupling) have been studied in \cite{Benn:1980ea,Obukhov:1983ri,Obukhov:1983mm,Saa:1993dv,Saa:1993eb,Saa:1996mt,Kirsch:2001ya,Mosna:2005mh}, and see \cite{Hehl:1994ue,Hehl:1999bt} for a discussion in more general spaces. For minimally coupled scalar fields, it is well known that the non-Riemannian pieces of the affine connection do not couple directly in the field equations. This result is not so clear for spin $1/2$ fields. Indeed it is not trivial  how to generalize the Dirac equation to non-Riemannian spacetimes\footnote{See \cite{Kirsch:2001ya} for a generalization of the Dirac equation to the framework of gauge theories of the affine group.} in a minimal coupling spirit, as the minimal coupling prescription applied to the Minkowskian spin 1/2 field equation (MCPF) gives a different result than when applied in the Minkowskian spin 1/2 Lagrangian (MCPL) \cite{Formiga}. Also recently, it has been claimed that the MCPF and MCPL give different dynamics for  matter fields in general in Riemann-Cartan space-times \cite{Chen:2019zhj}. Since in Riemannian space-times both prescriptions lead to the same dynamics, the aim of this work is trying to shed some light into the question posed in \cite{Formiga,Chen:2019zhj} of whether the MCPF or the MCPL is generally better suited to describe matter degrees of freedom in presence of torsion and/or non-metricity. To that end, we will we will try to shape or delimit the concept of minimal coupling for matter fields of spin 0, 1/2 and 1. We will also show how both MCPF and MCPL describe the same dynamics for scalar and vector fields if applied consistently, contrary to the claims of \cite{Chen:2019zhj}, and that the MCPF is not consistent with charge conservation for fermionic fields, thus addressing the question raised in \cite{Formiga}.\\
 
The structure of the paper is as follows. We will start in section \ref{sec:MinCoup} with a discussion on what does minimal coupling mean and why the naive recipe of replacing $\partial$ by $\na$ is not really a minimal coupling prescription when torsion and/or non-metricity do not vanish. In section \ref{sec:divopEL} we will derive a generalised form of the Euler-Lagrange (EL) equations for arbitrary spin matter fields in a space-time with generic torsion and non-metricity tensors. As a byproduct of the derivation of the Euler-Lagrange equations, we will see how non-metricity and torsion do not affect the functional form of conserved matter currents and charges, which have the same functional dependence on the matter fields as in the Riemannian case. Then, in section \ref{sec:matterfields }, we will see how applying a naive MCPF to the Klein-Gordon equation leads to field equations which are not equivalent to the generalised Euler-Lagrange equations for scalar and vector fields, while applying the MCPF as we defined leads exactly to the generalised EL equations. We will see how for spin 1/2 fields, even if the our MCPF is applied to the Dirac equation, the resulting equation is not equivalent to the one obtained by the generalised EL equations through our MCPL (if one uses the hermitian Minkowskian action). Indeed, while the generalised EL equations obtained through the MCPL are perfectly consistent to describe a spin 1/2 field with interactions that conserve an internal charge, the equations resulting from applying the MCPF leads to violation of this conservation. Thus, if charge conservation is assumed, the only valid minimal coupling prescription for spinor fields in presence of torsion  and/or non-metricity is the MCPL. We will also explictly show how non-metricity decouples from minimally coupled matter fields, although physical effects related to non-metricty could occur due to its non-trivial relation with the space-time metric (see \cite{Letter,Delhom:2019wir}). Finally, we add some concluding remarks in section \ref{sec:Final}.

\section{What is minimal coupling?} \label{sec:MinCoup}

It is now convenient to discuss the meaning of minimal coupling and the subtleties behind the naive prescription commonly used to implement it. Usually, the minimal coupling prescriptions (both MCPF and MCPL) are implemented by the following rule of thumb: \textit{Wherever you find a partial derivative $\partial$ or a Minkowski metric $\boldsymbol{\eta}$ in flat space-times, substitute them by the appropriate covariant derivative $\na$ and space-time metric $\bg$.} This sentence provides a recipe that works well in absence of torsion and/or non-metricity, although it is misleading about what a minimal coupling prescription  actually is, and it can lead to wrong results when torsion and/or non-metricity are non-vanishing. Indeed a more precise understanding of the minial coupling prescriptions can be achieved by arguing in the following way. As nicely argued in \cite{Wald:1984rg}, the operator $\partial$ acting on tensor fields is frame dependent, {\it i.e.} different coordinate systems have associated a different $\partial$ operator. Therefore $\partial$ will yield non-coordinate-invariant objects when applied over any tensor or spinor field. Hence, given that physical theories must be coordinate independent, we must not employ a particular $\partial$ operator in the construction of physical theories. Nonetheless, there are other derivative operators such as the covariant derivative $\nabla$, the exterior differential $\dif$, the co-differential $\boldsymbol{\delta}$, or suitable combinations of these that are the same operators in all coordinate systems, and therefore yield coordinate invariant objects when applied to tensor or spinor fields.\footnote{Note that $\dif$ and $\boldsymbol{\delta}$ are defined only for p-form fields, and not general tensor or spinor fields. Also note that $\dif$ is defined in any differentiable manifold without adding extra structure, $\bdel$ requires a volume form on the manifold, and $\na$ requires a connection on the manifold.} This makes $\partial$ inappropriate in the construction of physical theories even in flat space-times, since it will be a different operator in two non-inertially-related frames. Since usually field theories are formulated first in a Minkowskian context and in inertial coordinate frames, the symbol $\partial$ is commonly used, although one should recall that even in this case the derivative operators appearing in the the action and field equations are $\na$, $\dif$, $\bdel$ or appropriate combinations of them like the wave operator $\dif\bdel+\bdel\dif$. Thus, strictly speaking, an MCPF or MCPL prescription will be a minimal coupling prescription if it keeps track of the differential operators that are used in Minkowski space-time and uses the same operators in a general space-time. This can be summarised in the following rule of thumb:  {\it Substitute $\boldsymbol{\eta}$ by $\bg$ and use the same differential operators in Minkowski and in general space-times (in the sense of using} $\na$, $\dif$, $\bdel$ {\it or the appropriate combination that is also used in the Minkowskian case)}.\\

 The origin of the naive recipe for the minimal coupling prescriptions stems from its usefulness in Riemannian space-times, which can be argued as follows. For spinor fields there is no ambiguity in which is the operator employed since $\dif$ or $\bdel$ are not well defined in this case.  For scalar and vector fields (0- and 1-form fields respectively) the combinations\footnote{For the spin 1/2 field equation in Minkowski, the kinetic term usually written with $\partial$ should actually be understood as the $\na$ operator of the purely inertial connection (since $\dif$ or $\bdel$ are not well defined for spinors), while in bosonic equations the usual kinetic term is given by the wave operator $\Box_\varepsilon\equiv\dif\boldsymbol{\delta}_\varepsilon+\boldsymbol{\delta}_\varepsilon\dif\;\;$,  $\;\boldsymbol{\delta}_\varepsilon=\star_\varepsilon\dif\star_\varepsilon$ is the codiferential operator associated to the volume form $\varepsilon$, and $\star_\varepsilon$ stands for the Hodge operator associated to $\varepsilon$. $\Box_\varepsilon$ is also named d'Alambertian, Laplace-Beltrami operator, or form-Laplacian.  \label{footnotedalambertian}} in which $\partial$ appears in the respective actions are such that substituting $\partial$ by $\na$ gives the same result as using the $\dif$ operator when the space-time is Riemannian. Thus for the MCPL and in Riemannian space-times the naive $\partial\to\na$ gives the right answer. While this is also true for scalar fields if we aply the MCPF in Riemannian space-times, as we will see later, there is already a difference if we apply the MCPF as $\partial\to\na$ for vector (1-form) fields: the wave operator that governs the kinetic term of the field equations in Minkowski space-time can be written\footnote{Any metric defines a canonical volume form by the square root of its determinant. We will use the notation $g=|\text{det}(\bg)|$ and $\eta=|\text{det}(\boldsymbol{\eta})|$. Thus any subindex $g$ or $\eta$ in operators that depend on a volume form is interpreted as the operator that results dfrom replacing $\varepsilon$ by such volume form in the corresponding definition.\label{footnotevolume}} as $\Box_\eta=\eta^{\mu\nu}\partial_\mu\partial_\nu$, and thus it would be generalised to the operator $g^{\mu\nu}\na^\bg_\mu\na^\bg_\nu$ which gives a different result from the correct wave operator $\Boxg=\dif\bdel_g+\bdel_g\dif$ when applied to (1-form) fields (see section \ref{sec:vecfield} for details). Thus although the naive MCPF could lead to ambiguities in the vector field description, the naive MCPL prescription of replacing $\partial$ by $\na$ in the Minkowskian Lagrangians works perfectly well in Riemannian space-times. In the Riemannian case, this prescription is also known as {\it universal coupling} \cite{Will_Book}, and it is also worth to note that this universal coupling is a consistency requirement for any unitary and Lorentz invariant theory containing a massless spin 2 particle in its spectrum \cite{Weinberg:1965nx,Weinberg:1995mt}. \\
 
Nonetheless, as we will see later in detail, when we have other geometrical objects than the metric (such as an independent affine connection),  both MCPF and MCPL fail in being minimal coupling prescriptions for vector fields in non-Riemannian space-times if implemented through the naive rule $\na\to\partial$. This can be understood as follows: If following the naive MCPL we forget that the field-strength of the vector field is defined by $\boldsymbol{F}\equiv\dif \boldsymbol{A}$ and look only at its expression in some coordinate system $F_{\mu\nu}=2\partial_{[\mu}A_{\nu]}$, then the naive MCPL given by substituting $\partial\to\na$ in the vector action will lead us to a wrong definition of \textit{field-strength} $\tilde{F}_{\mu\nu}=2\na_{[\mu}A_{\nu]}$ which introduces an extra non-minimal coupling between torsion and the vector field, as opposed to just aplying the MCPL as we have defined, {\it i.e.} by keeping explicit track of the use of the $\dif$ operator in the vector Lagrangian. When trying to use the MCPL, the naive substitution of $\eta^{\mu\nu}\partial_\mu\partial_\nu$ by $g^{\mu\nu}\na_\mu\na_\nu$ instead of the use of the correct wave operator $\bdel\dif+\dif\bdel$ (which in Minkowski space is given by $\eta^{\mu\nu}\partial_\mu\partial_\nu$) introduces non-minimal couplings to the torsion and non-metricity tensor for both scalar and vector fields. For scalar fields, given that only first order derivatives appear in the action, there can be no naive MCPL since $\partial\Phi=\dif\Phi=\na\Phi$ by definition. However, the naive MCPF does give rise to non-minimal couplings between the scalar field and the torsion and non-metricity tensors.\\

\section{MCPL in non-Riemannian space-times: Euler-Lagrange equations and conserved currents}\label{sec:divopEL}

As it is common to work with actions and field equations written in terms of fields and their first covariant derivatives\footnote{Note that $\dif$ and $\delta$ can also be written in terms of $\na$ and the torsion tensor to recast the Lagrangian in this form.}, let us first derive the Euler-Lagrange (EL) equations in terms of the affine covariant derivative $\na$.  To that end, it will be useful to express the divergence operator associated with the metric in terms of the affine covariant derivative $\na$ employed in the action. This relation will allow us to employ Stokes' theorem in order to derive the generalised EL equations and the conserved currents and charges associated to continuous symmetries of the action.

\subsection{The divergence operator}

Let $\mathcal{M}$ be an n-dimensional smooth oriented manifold with volume n-form $\varepsilon=f dx^{\mu_1}\wedge...\wedge dx^{\mu_n}$ in some chart. The divergence operator associated to $\varepsilon$ acting on vector fields $A\in\mathfrak{X}(\mathcal{M})$ is the function $Div_\varepsilon:\mathfrak{X}(\mathcal{M})\rightarrow C^\infty(\mathcal{M})$ defined by \cite{Lee}\footnote{Some autors write the inner product $\boldsymbol{A}\lrcorner\varepsilon$ as $i_{\boldsymbol{A}} \varepsilon$}
 \begin{equation}\label{Divop}
 Div_\varepsilon(\boldsymbol{A})\varepsilon\equiv \dif(\boldsymbol{A}\lrcorner\varepsilon),
 \end{equation}
 which in a coordinate chart $x^\mu$ reads
\begin{equation}\label{Divopcoord}
Div_\varepsilon(\boldsymbol{A})=\frac{1}{f}\pa{\mu}(fA^\mu).
\end{equation}
This definition is completely general, as it only requires the differential structure of $\mathcal{M}$ and a general volume form defined on it\footnote{It turns out that this can also be written as $Div_\varepsilon(\boldsymbol{A})=\boldsymbol{\delta}_\varepsilon\boldsymbol{A}$, which can also be straightforwardly generalised to p-forms as $Div_\varepsilon(\boldsymbol{\Omega})=\boldsymbol{\delta}_\varepsilon\boldsymbol{\Omega}$ for any p-form $\boldsymbol{\Omega}$.}; neither a metric tensor $\bf{g}$ nor an affine structure $\bf{\Ga}$ on $\mathcal{M}$ are necessary.\\

The interest of this operator relies in that it satisfies a generalised divergence theorem, i.e. it relates a vector field defined in a volume $\mathcal{V}$ with its flux through the boundary $\partial\mathcal{V}$. As shown in \cite{Lee}, from Stokes' theorem one finds
\begin{equation}\label{Divtheogen}
\int_\mathcal{V}Div_\varepsilon(\boldsymbol{A})\varepsilon=\int_{\partial\mathcal{V}} \boldsymbol{A}\lrcorner\varepsilon;
\end{equation}
where $\mathcal{V}$ is the n-volume enclosed by $\partial\mathcal{V}$. This is the generalised divergence theorem for n-dimensional manifolds. \\

Now, if a metric structure $\bf{g}$ is introduced, there is a canonical choice for the volume n-form: $\varepsilon\equiv \dvg=g^{1/2}dx^{\mu_1}\wedge...\wedge dx^{\mu_n}$ (see footnote \ref{footnotevolume}). In this case, in $Lemma\; 16.30$ of \cite{Lee}, it is proven that\footnote{Here $\iota_S^*(\boldsymbol{A}\lrcorner \dvg)$ is the restriction of the (n-1)-form $\boldsymbol{A}\lrcorner \dvg$ to the boundary $\partial\mathcal{V}$ \cite{Lee}.}  $\iota_S^*(\boldsymbol{A}\lrcorner \dif V_\bg)=\bg(\boldsymbol{A},\boldsymbol{n})\dif V_{\tilde{g}}$, where $n$ is the unit normal to $\partial\mathcal{V}$ and $\dif V_{\tilde{g}}$ is the induced volume form on $\partial\mathcal{V}$ by $\bg$. Therefore, when a metric is present  (and is chosen to define the volume element),  the right hand side of \eqref{Divtheogen} can be interpreted as the flux \textit{normal} to the boundary enclosing $\mathcal{V}$. Thus the generalised divergence theorem \eqref{Divtheogen} can be seen as a generalised Gauss' law
\begin{equation}\label{Gaussgen}
\int_\mathcal{V}Div_g(\boldsymbol{A})\dif V_\bg=\int_{\partial\mathcal{V}} \bg(\boldsymbol{A},\boldsymbol{n})\dif V_{\tilde{\bg}}\;,
\end{equation}
 which relates the divergence of a vector field $A$ inside a closed volume $\mathcal{V}$ with the integration over $\partial\mathcal{V}$ of the component of $A$ normal to $\partial\mathcal{V}$. \\

Note that a metric $\bg$ also induces a canonical affine structure (or affine connection) $\bf{\Ga}=C(\bg)$ on $\mathcal{M}$ which is said to be compatible with it: the so called Levi-Civita connection of $\bg$. However, the affine structure on $\mathcal{M}$ needs not be compatible with \bg, and it is generally independent of it. Indeed, one can have a manifold with an affine structure but no metric structure. We say that a manifold where $\bg$ and $\bf{\Ga}$ are compatible is a \textit{Riemannian}\footnote{Note that matematicians call a Riemannian manifold one whith an affine structure compatible with the metric and with a metric of Riemannian signature. We will use Riemannian and non-Riemannian referring only to the (non-)compatibility of the connection and the metric and not to the signature of \bg, as it is often done in gravitational physics.} manifold $(\mathcal{M},\bg, \bf{C}(\bg))$; and one where $\bf{\Ga}$ and $\bg$ are independent is a \textit{non-Riemannian} manifold $(\mathcal{M},\bg, \bf{\Ga})$. In the following, we will generally work in a non-Riemannian manifold with the canonical volume element associated to its metric.\\

Any affine structure $\bf{\Ga}$ defines (and is defined by) a covariant derivative $\na$, which is completely specified by its connection symbols ${\Ga_{\mu\nu}}^\al$. The action of $\na$ on (the components of) an n-form \footnote{Remind that, since the space of n-forms in an n-dimensional manifols is of dimension 1, any n-form is proportional to the trivial ome $dx^1\wedge...\wedge dx^n$, and therefore is specified only by one component (the proportionality factor). This component is sometimes called a tensor density of weight $+1$.} $f$ and a vector field $A^\mu$ is given by
\begin{align}\label{DefCovDer}
\begin{split}
\nabla_\mu& f =(\pa{\mu}f-{\Ga_{\mu\al}}^\al f).\\
\nabla_\mu& A^\al=\pa{\mu}A^\al+{\Ga_{\mu\nu}}^\al A^\nu
\end{split}
\end{align}
Therefore using \eqref{Divopcoord} $Div(A)$ can also also be written as
\begin{equation}\label{Divtor}
Div_\varepsilon(\boldsymbol{A})=\frac{1}{f}\nabla_{\mu}(fA^\mu)-{S_{\mu\al}}^\al A^\mu,
\end{equation}
where ${S_{\mu\nu}}^\al\equiv-2{\Ga_{[\mu\nu]}}^\al$ is called the \textit{torsion} of the affine connection, and it identically vanishes in Riemannian manifolds. Note that \eqref{DefCovDer} and \eqref{Divtor} are true whether $\mathcal{M}$ has a metric structure or not.\\

In a non-Riemannian manifold, it is always possible to perform a decomposition of the connection symbols in the form ${\Gamma_{\mu\nu}}^{\alpha}={C_{\mu\nu}}^{\alpha}+{L_{\mu\nu}}^{\alpha}+{K_{\mu\nu}}^{\alpha}$; where
\begin{equation}\label{DecompConnSymb}
\begin{split}
&{C_{\mu\nu}}^{\alpha}\equiv\frac{1}{2}g^{\alpha\beta}\left(2\partial_{(\mu}g_{\nu)\beta}-\partial_\beta g_{\mu\nu}\right), \\
&{L_{\mu\nu}}^{\alpha}\equiv\frac{1}{2}\left(2Q_{(\mu\nu)}{}^\alpha-Q^\alpha{}_{\mu\nu}\right),\\
&{K_{\mu\nu}}^{\alpha}\equiv\frac{1}{2}g^{\alpha\beta}\left(2S_{(\mu|\beta|\nu)}-S_{\mu\nu\beta}\right) \ ;
\end{split}
\end{equation}
and where $Q_{\al\mu\nu}\equiv -\na_{\al}g\mn$ is the so-called \textit{non-metricity} tensor, which identically vanishes in Riemannian manifolds, ${L_{\mu\nu}}^{\alpha}$ is the distortion tensor, and ${K_{\mu\nu}}^{\alpha}$ is called contortion tensor (see e.g.  \cite{Ortin}). Here ${C_{\mu\nu}}^{\alpha}$ are the Christoffel symbols of $\bf{g}$, which are the connection symbols of the Levi-Civita connection ${\bf C}(\bg)$.  \\

From \eqref{DecompConnSymb} and the second equation in \eqref{DefCovDer} (identifying the volume form $f$ with\footnote{Note that this discussion could also be done in the tetrad formalism simply by writing the the volume in terms of the determinant of the tetrads instead of that of the metric without changing the results.} $g^{1/2}$), it is possible to show that $\na_\mu \lr{g^{1/2}}=-\frac{1}{2}{Q_{\mu\al}}^\al g^{1/2}$. Using this identity, from \eqref{Divtor} one finds the following relation
\begin{equation}\label{Divgen}
Div_g(\boldsymbol{A})=\na_\mu A^\mu-\lr{\frac{1}{2}{Q_{\mu\al}}^\al+{S_{\mu\al}}^\al}A^\mu,
\end{equation}
which is the sought relation between the divergence operator, the covariant derivative, and the non-metricity and torsion tensors of $\mathcal{M}$. For Riemannian manifolds, where $Q_{\mu\nu\al}$ and ${S_{\mu\nu}}^\al$ vanish by definition, \eqref{Divgen} reduces to the usual expression $Div(A)=\na^\bg_\mu A^\mu$ as it must be, where $\nabla^\bg_\mu$ is the covariant derivative associated to the Levi-Civita conection of the metric $\bg$.\\

 Notice also that from the definition of ${C_{\mu\nu}}^\al$ in \eqref{DecompConnSymb} it is possible to get the following useful relation
\begin{equation}\label{IdentityderivLC}
g^{-1/2}\pa{\mu}(g^{1/2}A^\mu)=\pa{\mu}A^\mu+{C_{\nu\mu}}^\nu A^\mu\equiv\na^\bg_\mu A^\mu.
\end{equation}
Since from the definition of divergence operator we have \eqref{Divopcoord}, we have also generally that
\begin{equation}\label{DivLevi}
Div_g(\boldsymbol{A})=\na^\bg_\mu A^\mu;
\end{equation}
 Therefore, we end up with the result that in any manifold where the volume element is given by the metric, no matter what the affine structure is, the identity \eqref{DivLevi} holds, which can be summed up in the following equation:
\begin{equation}\label{Divgen2}
Div_g(\boldsymbol{A})=\na_\mu A^\mu-\lr{\frac{1}{2}{Q_{\mu\al}}^\al+{S_{\mu\al}}^\al}A^\mu=\na^\bg_\mu A^\mu
\end{equation}
for any $(\mathcal{M},\bg, \na)$. From the above discussion one can infer that, in a non-Riemannian manifold, the relation between the divergence operator (given by the metric structure) and the covariant derivative of the manifold is not the same as in a Riemannian manifold due to the fact that the affine structure has no relation with the volume form, and torsion and non-metricity have to be taken into account. On the other hand, in a Riemanian manifold, given that $\na=\na^\bg$, the affine structure and the volume element are indeed related, which translates into a direct relation between the divergence operator and covariant derivative of the manifold. 
\begin{widetext}
\subsection{Euler-Lagrange equations of a Minimally coupled theory in non-Riemannian spacetimes}

Clarifying the relation between the divergence operator and the affine structure will be useful to derive the generalised EL equations for any minimally coupled matter Lagrangian (although it can also be done without using the above relations). To do so, we start by applying the MCPL to the usual Minkowski Lagrangian as explained in section \ref{sec:MinCoup}, so that we start with a functional $\lag(\Psi_i,\partial\Psi_i)$ that defines the following action: 
\begin{equation}\label{matteraction}
S_m\lrsq{\Psi_i,\na_\mu\Psi_i}=\displaystyle \int_\mathcal{V}\dvg\mathcal{L}[\Psi_i,\na_\mu\Psi_i] \ ,  \qquad i=1,...,N \ ;
\end{equation}
where $\mathcal{L}[\Psi_i,\na_\mu\Psi_i]$ is a scalar and $\{\Psi_i\}$ is the collection of matter fields. Let us point out that, if $\dif$ instead of $\na$ is employed in the construction of $S_m$, as is the case for vector fields, we can always re-write them as covariant derivatives plus some extra terms proportional to the torsion tensor. Therefore  \eqref{matteraction} is rather general, since any minimally coupled matter action can be recast in such form.\\

 The field equations are obtained, as usual, by applying the extremal action principle and then solving the variational problem $\delta S_m=0$ for some arbitrary variations of the matter fields $\delta\Psi_i$ vanishing at the boundary of $\mathcal{V}$. Notice though that given variation of the field $\delta\Psi_i$ naturally introduces also a variation in its partial derivartive $\delta(\partial\Psi_i)$, and the variational problems that one is used to solve are in terms of the field variables $\{\Psi_i, \partial\Psi_i\}$. Since by definition $\na_\mu\Psi_i\equiv\partial_\mu\Psi_i-\Upsilon^{i}_\mu\Psi_i$, where $\Upsilon^{i}_\mu$ are the connection coefficients in the representation of $\Psi_i$, we can re-write the above action in terms of the independent variables $\{\Psi_i,\partial_\mu\Psi_i\}$ and proceed with standard variational methods. By explicitly substituting $\na_\mu\Psi_i$ by its expression in terms of $\{\Psi_i,\partial_\mu\Psi_i\}$ we can recast \eqref{matteraction} as a function of the variables $\{\Psi_i,\partial_\mu\Psi_i\}$ instead of $\{\Psi_i,\na_\mu\Psi_i\}$, thus having    
\begin{equation}\label{matteractionsubst}
\tilde S_m\lrsq{\Psi_i,\pa{\mu}\Psi_i}=\displaystyle \int_\mathcal{V}\dvg\mathcal{L}[\Psi_i,\na_{\mu}\Psi_i(\Psi_i,\pa{\mu}\Psi_i)]=\displaystyle \int_\mathcal{V}\dvg\mathcal{L}[\Psi_i,\pa{\mu}\Psi_i-\Upsilon_\mu^i\Psi_i],
\end{equation}
where the tilde here emphasizes that the functional form of $S_m$ changes when we consider it as a functional of the new variables. Now we can employ the standard methods of variational calculus to solve $\delta\tilde S_m=0$ for an arbitrary variation $\delta\Psi_i$ that  vanishes at the boundary of $\mathcal{V}$, which leads to
\begin{align}\label{deltatildeS3}
\delta \tilde S_m=\displaystyle \int_\mathcal{V}\dvg \lrsq{\lr{\frac{\pa{}\lag}{\pa{}\Psi_i}-\frac{\partial\lag}{\partial(\na_\mu\Psi_i)}\Upsilon^i_\mu}\delta\Psi_i+\lr{\frac{\pa{}\lag}{\pa{}(\na_\mu\Psi_i)}}\delta(\partial_\mu\Psi_i)}=0,
\end{align}
where we have used that
\begin{equation}
\frac{\partial(\na_\mu\Psi_i)}{\partial\Psi_i}=-\Upsilon_\mu^i\qquad , \qquad\frac{\partial(\na_\nu\Psi_i)}{\partial(\partial_\mu\Psi_i)}=\delta^\mu{}_\nu.
\end{equation}
We now want to invert the change of variables by writing $\partial_\mu\Psi_i$ as a function of the old independent variables $\{\Psi_i, \na_\mu\Psi_i\}$, which is by definition given by $\partial_\mu\Psi_i=\na_\mu\Psi_i+\Upsilon_\mu^i\Psi_i$. Then, we can write an arbitrary variation of $\partial_\mu\Psi_i$ as a function of an arbitrary variation of the old variables: $\delta(\pa{\mu}\tilde\Psi_i)=\delta(\na_\mu\Psi_i)+\Upsilon_\mu^i\delta\Psi_i$. Plugging this expression in the above equation, we end up with\\

\begin{align}\label{deltaSprimary}
\delta S_m=\displaystyle \int_\mathcal{V}\dvg \lrsq{\frac{\pa{}\lag}{\pa{}\Psi_i}\delta\Psi_i+\frac{\pa{}\lag}{\pa{}(\na_\mu\Psi_i)}\delta(\na_\mu\Psi_i)}=0,
\end{align}
which by means of \eqref{Divgen} can be recast into

\begin{align}\label{deltaScomplete}
\delta S_m=\displaystyle\int_\mathcal{V} \dvg \lrsq{\frac{\pa{}\lag}{\pa{}\Psi_i}-\na_\mu\lr{\frac{\pa{}\lag}{\pa{}(\na_\mu\Psi_i)}}+\lr{\frac{1}{2}{Q_{\mu\al}}^\al+{S_{\mu\al}}^\al}\lr{\frac{\pa{}\lag}{\pa{}(\na_\mu\Psi_i)}}}\delta\Psi_i+\displaystyle\int_\mathcal{V} \dvg Div_g\lr{\frac{\pa{}\lag}{\pa{}(\na_\mu\Psi_i)}\delta\Psi_i}=0.
\end{align}
Since the last term is a boundary term by the generalised Gauss' law \eqref{Gaussgen}, it vanishes for variations $\delta\tilde\Psi_i=\delta \Psi_i$ vanishing at the boundary of $\mathcal{V}$. Thus, the above equation leads to to the non-Riemannian version of the covariant Euler-Lagrange equations, which reads
 \begin{equation}\label{eom}
 \frac{\pa{}\lag}{\pa{}\Psi_i}-\na_\mu\lr{\frac{\pa{}\lag}{\pa{}(\na_\mu\Psi_i)}}+\lr{\frac{1}{2}{Q_{\mu\al}}^\al+{S_{\mu\al}}^\al}\lr{\frac{\pa{}\lag}{\pa{}(\na_\mu\Psi_i)}}=0.
\end{equation}
Note that while in the Riemannian limit we recover the usual covariant Euler-Lagrange equations, in the general case there are, apparently, explicit couplings between the non-metricity and torsion tensors and the matter fields. However, these apparent couplings are indeed compensated by taking into account that the covariant derivative of the second term in \eqref{eom} is not the one associated to the Levi-Civita connection of $\bg$. To show this, we can use \eqref{DecompConnSymb} and split the covariant derivative in front of the second term of \eqref{eom}, thus re-writing the non-Riemannian Euler-Lagrange equations \eqref{eom} as
 \begin{equation}\label{eom2}
 \frac{\pa{}\lag}{\pa{}\Psi_i}-\na^\bg_\mu\lr{\frac{\pa{}\lag}{\pa{}(\na_\mu\Psi_i)}}-\lr{\frac{\pa{}\lag}{\pa{}(\na_\mu\Psi_i)}}(\Upsilon^{i}_{NR})_\mu=0,
\end{equation}
where $(\Upsilon^{i}_{NR})_\mu$ is the non-Riemannian part of the connection in the representation corresponding to $\Psi_i$ ({\it i.e.} the piece of $\Upsilon^i_\mu$ that features torsion and non-metricity) and $\na^\bg$ is the Riemannian covariant derivative. Here we can see how the explicit couplings to non-metricity and torsion that are present in \eqref{eom} actually cancel out and do not contribute to the dynamics, and the only possible source of non-Riemannian couplings comes from the $(\Upsilon^{i}_{NR})_\mu$, and from $\na_\mu\Psi_i$ in the derivatives of the Lagrangian. This could not have been otherwise because of the following reason: Since in the action \eqref{matteraction} the only possible non-Riemannian couplings appear through the $(\Upsilon^i_{NR})_\mu\Psi_i$ term in the $\na_\mu\Psi_i$ variables, the only non-Riemannian terms  that can show up in the field equations will also enter through $(\Upsilon^i_{NR})_\mu$. From this result follows the conclusion that whether a minimally coupled (in the sense of MCPL) matter field couples or not to the non-Riemannian features of a general space-time depends only on the form of the connection in its corresponding spin representation. As we will see later, in the case of spin 0 and 1 fields, all the explicit couplings between non-metricity and torsion dissappear form the field equations if the minimal coupling prescriptions that we have outlined are applied correctly. On the other hand, in the case of spin 1/2 fields, a residual interaction with the totally antisymmetric part of the torsion tensor remains for the MCPL, and a more complicated coupling to the traces of torsion and non-metricity occurs for the MCPF. This coupling will be seen to source a charge violating current.

\end{widetext}

\subsection{Conserved currents and charges }\label{sec:currents}

As a by-product of the derivation of \eqref{eom}, we can also investigate whether the functional form of the Noether currents/charges associated to matter fields will be modified by non-metricity or torsion corrections. For completeness, let us briefly explain the geometrical meaning of a conserved current. By definition, a conserved vector current over a spacetime $\mathcal{M}$ equipped with a volume form $\varepsilon$ is a vector field ${\boldsymbol{J}}\in\mathfrak{X}(\mathcal{M})$ that satisfies 
\begin{equation}\label{defconscurrent}
Div_\varepsilon({\boldsymbol{J}})=0
\end{equation}over $\mathcal{M}$. This definition is valid for Riemannian as well as for non-Riemannian spacetimes, and it has an intuitive geometrical meaning as will be clarified below. Notice that provided that the volume form is chosen to be the one given by the metric, in Riemanninan spacetimes condition \eqref{defconscurrent} is equivalent to $\na_\mu J^\mu =0$, while in non-Riemannian spacetimes it cannot be written simply as $\na_\mu J^\mu =0$. The importance of the condition \eqref{defconscurrent} relies in that it allows one to define a scalar quantity ${\bf{Q_J}}$ on every spatial hypersurface of a Cauchy foliation of the given space-time such that ${\bf{Q_J}}$ is invariant under time-translation (i.e. change of spatial hypersurface). This is why a vector satisfying $Div_\varepsilon(\boldsymbol{J})=0$ is called a \textit{conserved current}, and ${\bf{Q_J}}$ is its associated \textit{conserved charge}.\\

\begin{figure}\label{fig:ConservedCharge}
\center\includegraphics[scale=0.2]{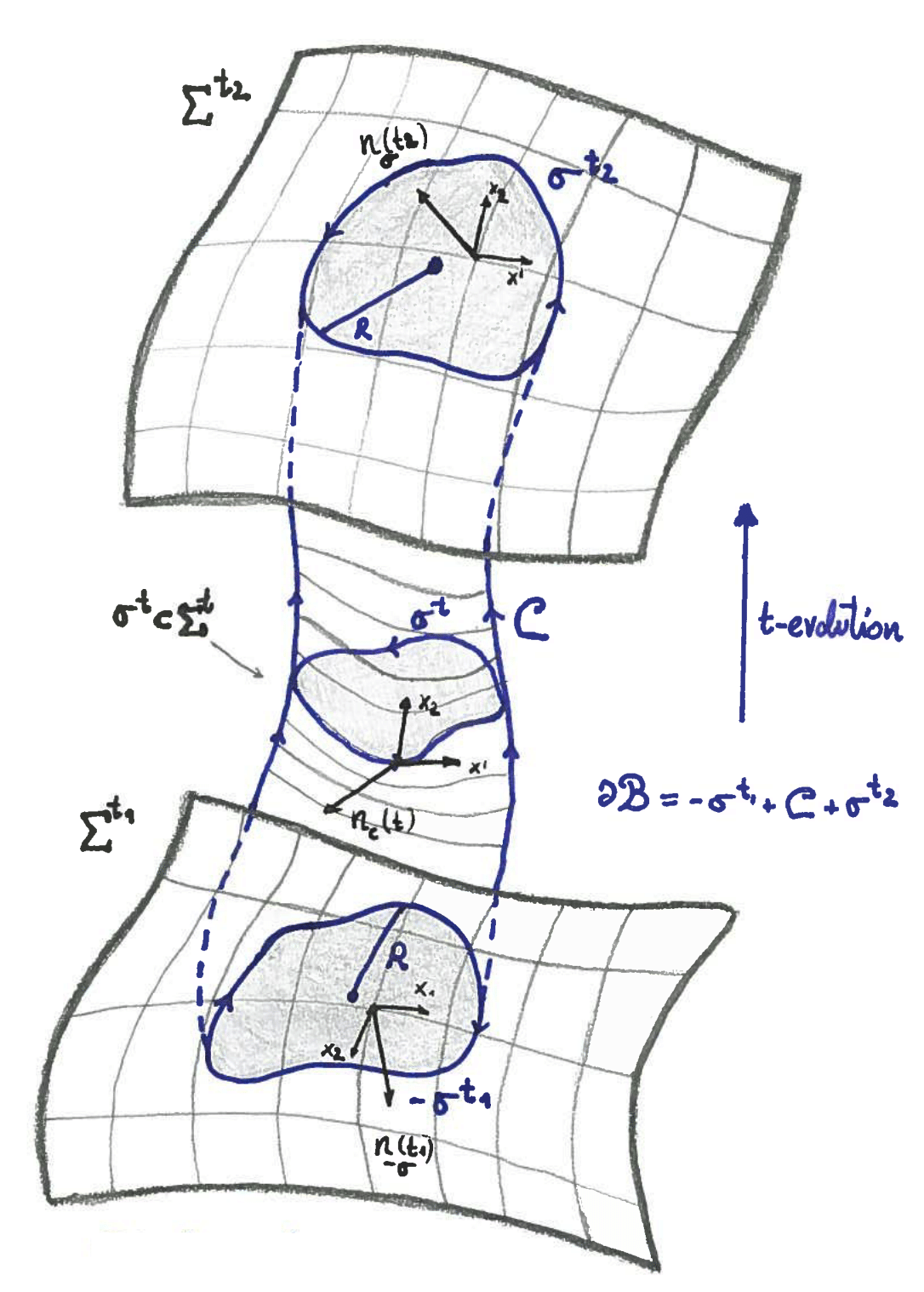}
\caption{Illustration of the different parts of $\cB$ and $\partial\cB$. The black oriented vector basis define the orientation of $\partial\cB$.}
\end{figure}

 In order to clarify the geometrical meaning of the condition \eqref{defconscurrent}, let us first precisely formulate the existence of such conserved charge ${\bf Q_J}$ in a space-time with volume form given by the metric. Consider coordinates $(x^0=t,x^i)$ and a foliation of spacetime given by the one-parameter family of space-like 3-surfaces ${\Sigma^t}$ normal to $\pa{t}$. Consider also a 3-ball $\sigma^t$ defined in every $\Sigma^t$ by $(x^i x_i)^{1/2}\leq R$, with $R$ an arbitrarily big constant. Define the closed 4-volume $\mathcal{B}(t_1,t_2)$ enclosed by $\sigma^{t_1}$, $\sigma^{t_2}$ and $\cC$; where $\cC$ is the union of the boundaries of each $\sigma^t$ for $t\in(t_1, t_2)$ (see fig. \ref{fig:ConservedCharge} for clarification). Any vector field $\boldsymbol{J}$ defines a charge ${{\bf Q}^t_{\bf J}}$ at each $\Sigma^t$ given by
\begin{equation}\label{ChargeDef}
{\bf Q}^t_{\bf J}=\underset{R\rightarrow\infty}{lim}\int_{\sigma^t}J^t \dif V_{\tilde{g}} \ ,
\end{equation}
where $J^t=\bg({\bf J},\pa{t})$, $\pa{t}$ is the unit normal to $\sigma^t$ and $\dif V_{\tilde{g}}$  is the volume form induced on $\Sigma^t$ by $\dvg$. Using Gauss' law \eqref{Gaussgen}, decomposing $\partial\cB$ as\footnote{The sign infront of $\sigma^{t_1}$ is required for $\partial\mathcal{B}(t_1,t_2)$ to have the standard induced orientation from $\mathcal{V}(t_1,t_2)$ \cite{Lee}.} $\partial\mathcal{B}(t_1,t_2)=(-\sigma^{t_1}+\mathcal{C}+\sigma^{t_2})$, and for configurations of $\Psi$ such that $J^t$ vanishes quickly enough at spatial infinity\footnote{The precise requirement is that $\Psi$ vanish quiclky enough with increasing $R$ so that the integal over $\cC$ vanishes when $R\rightarrow\infty$.} we find
 
\begin{equation}\label{ChargeDiv}
\int_{\mathcal{B}}Div_g({\bf J})\dvg={\bf Q}^{t_2}_{\bf J}-{\bf Q}^{t_1}_{\bf J}.
\end{equation}

This is valid for any value of $t_1$ and $t_2$; and in particular, for infinitesimally small values of $\delta t=t_2-t_1$ we find $$\mathcal{L}_t{\bf Q}_{\bf J}|_{t=t_1}\delta t=\lr{{\bf Q}_{\bf J}^{t_2}-{\bf Q}_{\bf J}^{t_1}}=\int_{\mathcal{B}}Div_g({\bf J})\dvg.$$ Therefore a charge defined by a conserved vector current remains constant under time-evolution, i.e. we say it is \textit{conserved}. The arguments within this section are independent of the choice of connection ${\bf \Gamma}$, which remarks the importance of the condition $Div_g({\bf J})=0$ instead of any condition involving any covariant derivative, which points out that the expression \textit{``conserved with respect to a connection''}, usually found in the literature, can be misleading. Indeed, these arguments depend only on the choice of the volume form, and do not depend on the metric itself. Therefore, a more correct expression could be \textit{conserved with respect to a volume form, or a metric}. \\

From the above discussion, it is now clear that the condition \eqref{defconscurrent} implies that the change in the amount of charge ${\bf Q}_{\bf J}$ enclosed in the 3-volume $\sigma^{t_1}$ is given exactly by the flux of the current ${\bf J}$ through the 4-volume $\mathcal{B}$, i.e. the amount of charge that exits $\sigma^t_1$ in the time interval $t_2-t_1$. Therefore, total charge cannot be created or destroyed if condition \eqref{defconscurrent} holds.\\

We can now proceed to investigate wether the functional form of Noether matter currents should be modified to account for non-Riemannian features. Noether currents have proven to be very useful tools for analyzing and extracting physical information of field theories. They have been used in the context of extended gravity by the use of the so called Noether symetry approach (see for instance \cite{DeLaurentis1,DeLaurentis2,DeLaurentis3,DeLaurentis4}); and they lie at the very definition of physical charges. Noether currents are defined from a symmetry of an action, and are conserved if the corresponding symmetry is realised. Let us try to understand whether torsion or non-metricity  play any role in the physics associated to Noether currents and charges\\

Given an action like \eqref{matteraction} and a transformation of the matter fields $\delta\Psi_i$ which is a continuous symmetry of that action, we can work out the functional form for the Noether current associated to this symmetry by the following argument. The first term in \eqref{deltaScomplete} vanishes for fields satisfying their equations of motion \eqref{eom}. Since $\delta\Psi_i$ is associated to a continuous symmetry of the action \eqref{matteraction}, we have $\delta S_m=0$, hence, the second term in \eqref{deltaScomplete} must also vanish, yielding $$Div\lr{\frac{\pa{}\lag}{\pa{}(\na_\mu\Psi_i)}\delta\Psi_i}=0.$$ Therefore, as stated by the Noether theorem \cite{Noether}, a continuous symmetry of the matter action implies the existence of an associated conserved current ${\bf J}\in\mathfrak{X}(\mathcal{M})$ defined by
\begin{equation}\label{Noethercurrent}
J^\mu=\frac{\pa{}\lag}{\pa{}(\na_\mu\Psi_i)}\delta \Psi_i,
\end{equation}
and its corresponding conserved charge given by \eqref{ChargeDef}. A current defined from a Lagrangian as in \eqref{Noethercurrent} is called Noether current, and the corresponding charge is called  Noether charge. Physical charges are the Noether charges associated to some continuous global symmetry of the matter action. For instance, the electric and color charges are associated to the global $U(1)_{EM}$ and $SU(3)_C$ symmetries of the Standard Model action. The above equation \eqref{Noethercurrent} shows that Noether currents (and charges) have the same functional form in Riemannian and non-Riemannian spacetimes in terms of the covariant derivative of the matter fields. In other words, non-metricity and torsion do not change the functional form of the currents and charges. However there could in principle be implicit corrections entering through $\na_\mu\Psi_i$. However, we will see that in the case of scalar and spinor fields, there are no corrections due to the fact that $\na_\mu\Psi_i$ enters linearly in the action. By using the above results, we will also show later that the requirement of charge conservation in non-Riemannian space-times will serve as a discriminator between the MCPF and MCPL for spin 0 and 1/2 fields.

\section{Minimally coupled matter fields in non-Riemannian geometries}\label{sec:matterfields }

 In this section we will employ the results previously derived and work out the examples of minimally coupled scalar, spinor and vector fields, focusing on the viability or distinguishability of the MCPF and MCPL prescriptions, both in its naive version and in the version that we have defined. Also we will show whether they couple minimally to non-metricity and torsion for the different versions of each prescription.  As it will become apparent later, the explicit couplings between minimally coupled matter fields and non-metricity disappear from the  field equations in all the examples when the MCPF and MCPL as we have defined are applied. However it could still induce non-trivial modifications to the dynamics of matter fields through their coupling to the metric: Since non-metricity is a tensor related to both, metric and affine connection, a coupling to the metric can encode some non-metric effects, as suggested in \cite{Letter,Delhom:2019wir}. Also, if our MCPF and MCPL are applied, torsion does not couple either to minimally coupled scalar and vector fields, although its totally antisymmetric part does couple to minimally coupled spin 1/2 fields. On the other hand, we will see how the naive MCPF and MCPL generally lead to non-minimal couplings between matter fields, and the torsion and non-metricity tensors.

\subsection{The minimally coupled scalar field}

The action of a complex scalar field can be written in Minkowski space-time with any of the operators $\partial$, $\dif$ and $\nabla$, since they are all the same when acting on scalar fields. However, the wave operator $\Box_g$ of the Klein-Gordon equation is defined only for p-forms, so that for simplicity we will view $\Phi$ as a 0-form. However, in order to be able to use the generalised EL equations \eqref{eom}, we will write the action in terms of $\na_\mu\Phi$. Thus, by applying the MCPL, the action of a complex scalar field in a general background geometry reads as 
\begin{equation}\label{scalaraction}
S_\Phi=\displaystyle \int_\mathcal{V}\dvg \lrsq{g^{\mu\nu}\na_\mu\Phi^\dagger\na_\nu\Phi-m^2\Phi^\dagger\Phi}.
\end{equation}
The field equations that follow after using \eqref{eom} are
\begin{align}\label{eomscalarcomplex}
\begin{split}
&\Box_{\bf\Ga}\Phi+\Sigma^\mu\na_\mu\Phi+m^2\Phi=0,\\
&\Box_{\bf\Ga}\Phi^\dagger+\Sigma^\mu\na_\mu\Phi^\dagger+m^2\Phi^\dagger=0,\\
\end{split}
\end{align}
where we have defined $\Box_{\bf\Ga}\Phi\equiv g^{\mu\nu}\na_\mu\na_\nu\Phi$ and the non-Riemannian current
\begin{equation}\label{NRcurrent}
\Sigma^\mu\equiv{{Q_\al}^\al}^\mu-\frac{1}{2}{Q^\mu{}_{\al}}^\al-{S^\mu{}_{\al}}^\al.
\end{equation}
This is in apparent contradiction with the expectations that neither non-metricity, nor torsion couple to a minimally coupled scalar field, given that they do not appear in the action \eqref{scalaraction}. Nonetheless, by using again the decomposition of the affine connection \eqref{DecompConnSymb} it follows that
 \begin{equation}\label{DecompBox}
 \Box_{\bf\Ga}\Phi=\Box_g\Phi-g^{\mu\nu}\lr{{L_{\mu\nu}}^\al+{K_{\mu\nu}}^\al}\na_{\al}\phi=\Box_g\Phi-\Sigma^\mu\na_\mu\Phi,
 \end{equation}
 with $\Box_g=\dif\bdel_g+\bdel_g\dif$ (see footnote \ref{footnotedalambertian}), and where the same identity holds for $\Phi^\dagger$. Using \eqref{DecompBox}, the above field equations \eqref{eomscalarcomplex} can be re-written as 
\begin{align}\label{eomscalar}
\begin{split}
&(\Boxg+m^2)\Phi=0,\\
&\Phi^\dagger(\overleftarrow{\Box}_g+m^2)=0,
\end{split}
\end{align}
which, as expected, show that minimally coupled scalar fields do not couple to torsion and non-metricity explicitly. A more clever way to find this result is by noticing that, since for scalar fields $\na_\mu\Phi=\pa{\mu}\Phi$, we have that the connection coefficients for the scalar representation vanish, i.e. $\Upsilon^\Phi_\mu=0$, and in particular $(\Upsilon_{NR})^\Phi_\mu=0$ for scalar fields. Thus the scalar field equations as written in \eqref{eom2} are directly given by
\begin{equation}\label{ScalarEulerLagrangeDIv}
 \frac{\pa{}\lag}{\pa{}\Phi}-\na_\mu^\bg\lr{\frac{\pa{}\lag}{\pa{}(\na_\mu\Phi)}}=0
\end{equation}
or
\begin{equation}
\frac{\pa{}\lag}{\pa{}\Phi}-Div\lr{\frac{\pa{}\lag}{\pa{}(\na_\mu\Psi_i)}}=0 ,
\end{equation}
which lead directly to \eqref{eomscalar} without passing through \eqref{eomscalarcomplex}. This trick does not work, however, for fields of arbitrary spin, since the quantity $(\Upsilon_{NR})^i_\mu$ need not vanish in other representations, or seen it otherwise, $\partial\lag/(\partial\na_\mu\Psi)$ will no longer be a vector field and then the last terms of  \eqref{eom} are not the divergence operator acting on $\partial\lag/(\partial\na_\mu\Psi)$.\\

Now let us analyse what happens if we apply the MCPF as described in section \ref{sec:MinCoup} directly to the field equations. The scalar field equations in Minkowski space-time written in a frame-independent way are: 
\begin{align}\label{eomscalarcomplexMCPF}
\begin{split}
&(\Box_\eta+m^2)\Phi=0,\\
&\Phi^\dagger(\overleftarrow{\Box}_\eta+m^2)=0.\\
\end{split}
\end{align}

where $\Box_\eta=\dif\bdel_\eta+\bdel_\eta\dif$ (see footnote \ref{footnotedalambertian}). The MCPF as we have defined it tells us to replace $\Box_\eta$ by $\Box_g$, leading to the same field equations as our MCPL, given by $\eqref{eomscalar}$. Therefore, we conclude that, if applied as we have defined them, both MCPF and MCPL give the same results for scalar fields. The naive MCPL, in this case, will also be consistent with the above equations, since $\na\Phi=\dif\Phi$ for any affine connection, and there are only first order derivatives of the scalar field in the action \eqref{scalaraction}. However, if we had applied the naive MCPF in the scalar field equation, substituting $\Box_\eta=\eta^{\mu\nu} \partial_\mu\partial_\nu$ by $\Box_{\bf\Ga}=g^{\mu\nu}\na_\mu\na_\nu\Phi$, we would have arrived to the field equations

\begin{align}\label{eomscalarcomplexMCPFwrong}
\begin{split}
&(\Box_{\bf\Ga}+m^2)\Phi=0,\\
&\Phi^\dagger(\overleftarrow{\Box}_{\bf\Ga}+m^2)=0.\\
\end{split}
\end{align}

 which using \eqref{DecompBox} can be written as
\begin{align}\label{eomscalarcomplexMCPFexplicit}
\begin{split}
&\lrsq{\Boxg-\Sigma^\mu\na^\bg_\mu+m^2}\Phi=0\\
&\Phi^\dagger\lrsq{\overleftarrow{\Box}_g-\Sigma^\mu\overleftarrow{\na}^\bg_\mu+m^2}=0,
\end{split}
\end{align}
where $\Sigma^\mu$ is the non-Riemannian current \eqref{NRcurrent}. It is clear that, in the case of applying the naive MCPF leads to a non-minimal coupling between the scalar field and the torsion and non-metricity tensors through the current $\Sigma^\mu$, and therefore this cannot be regarded as a  minimal coupling prescription in the sense defined in section \ref{sec:MinCoup}.\\

 Let us now see what happens with charge conservation if we apply MCPL as we have defined it or the naive MCPF. In Riemannian space-times both the naive MCPF and our version of  the MCPL give rise to identical dynamics for scalar fields. Since the action obtained by the MCPL for a complex scalar field has a global $U(1)$ symmetry\footnote{The symmetry group need not be $U(1)$, it can be any Lie group if the scalar is a multiplet in the corresponoding representation}, it will have associated a conserved current given by \eqref{Noethercurrent}. Hence, the scalar field that evolves as given by our MCPL as well as the naive MCPF will both obey charge conservation. However, the fact that in non-Riemannian space-times the dynamics for scalar fields is different for our MCPL and the naive MCPF spoils the above argument. In this case, the MCPL still leads to an action with a global $U(1)$ symmetry \eqref{scalaraction}, and therefore the usual conserved scalar current can be derived from the scalar action \eqref{scalaraction} by applying the Noether current formula \eqref{Noethercurrent}, thus obtaining
 \begin{equation}\label{scalarcurrent}
 J^\mu_\Phi=i\lr{\Phi^\dagger(\na^\mu \Phi)-(\na^\mu \Phi^\dagger)\Phi},
 \end{equation}
which is still conserved for both the MCPL and MCPF as we have defined them, as can be seen by taking the divergence of ${\bf J}_\Phi$ and using \eqref{eomscalar}. However, if one follows the naive MCPF, since it differs in general from that with dynamics given by the MCPL in presence of torsion and/or non-metricity, the above current will generally not be conserved. Indeed, by taking the divergence of \eqref{scalarcurrent} and using the naive MCPF field equations \eqref{eomscalarcomplexMCPFexplicit}, we arrive to
\begin{equation}
Div_g({\bf J}_\Phi)\big|_{MCPF}=\Sigma_\mu J^\mu_\Phi,
\end{equation}
which shows that for a complex scalar field described by the naive MCPF, the non-Riemannian current that couples it to non-metricity and torsion spoils the potential $U(1)$ symmetry of the scalar field\footnote{There is an exception if both torsion and the non-metricity are traceless, since the non-Riemannian current $\Sigma^\mu$ vanishes in this case and both MCPF and MCPL as we have defined them are equivalent to the naive MCPF, recovering the $U(1)$ symmetry for the naive MCPF in this particular case.}. This argument applies in a straightforward manner for any complex scalar fields which are in a given representation of some Lie group (not only $U(1)$). Given that the Standard Model (SM) requires for its construction that the Higgs field Lagrangian be globally invariant under $SU(2)\times U(1)$, the naive MCPF prescription would enter in contradiction with the SM in presence of a (strong enough) torsion and/or non-metricity background. Since we do not have experimental data on how do matter fields behave in the presence of non-Riemannian features, there is no way to favour MCPF or MCPL over other non-minimal coupling prescriptions. However, one expects the SM to work perfectly well if tiny non-metricity and/or torsion corrections are included, therefore pushing the naive MCPF to an uncomfortable corner when applied to complex scalar fields. Note that the breaking of charge conservation will happen in general for any non-minimal coupling prescription that is applied directly to the field equations, since that would generically break the $U(1)$ invariance enjoyed by the scalar action.

\subsection{The minimally coupled spin $1/2$ field}

Let us now examinate the case of spin 1/2 fields. Contrary to the spin 0 case, the only derivative operator that acts on spinors and is covariant under arbitrary changes of basis is $\na$ (see e.g. \cite{Dabrowski:2013spinors,Dabrowski:1986cf}). Thus we conclude that the Minkowskian action is written, in a frame independent way, in terms of the covariant derivative $\na$ of the purely inertial connection (which satisfies $\partial=\na$ in any inertial frame). Thus, applying the MCPL as stated in section \ref{sec:MinCoup} for the two different actions that are commonly used in the literature to describe spinor fields in Minkowski space-times we are lead to
\begin{align}
&S_\psi=\displaystyle\int_{\mathcal{V}}\dvg\lrsq{\frac{i}{2}\lr{\bpsi\ga^\mu(\na_\mu\psi)-(\na_\mu\bpsi)\ga^\mu\psi}-\bpsi m\psi}\label{SpinorActionH}, \\
&\tilde{S}_\psi=\displaystyle\int_{\mathcal{V}}\dvg\lrsq{\bpsi\lr{i\ga^\mu\na_\mu-m}\psi},\label{SpinorActionWrong}
\end{align}
 where $\na_\mu\psi=\pa{\mu}\psi-\Upsilon^\psi_\mu\psi$ and $\na_\mu\bpsi=\pa{\mu}\bpsi+\bpsi\Upsilon^\psi_\mu$. There is a subtlety here:  we must first specify a form for the spinor connection $\Upsilon^\psi_\mu$. Generally, the spinor connection is defined in terms of the spin conection $\omega_\mu{}^{ab}$ as 
\begin{equation}\label{CanonicalSpinorConnection}
\Upsilon^\psi_\mu=\frac{1}{2}(\omega^{\bf\Ga})_\mu{}^{ab}\sigma_{ab}.
\end{equation}
 where $\sigma_{ab}=\frac{1}{4}\lrsq{\ga_a,\ga_b}$ are the generators of the Lorentz group in the spin representation \cite{Ortin,Hurley}. Thus $\Upsilon^\psi_\mu$ is completely specified once a choice for spin connection is made. In Riemannian space-times, there is a canonical lift of the Levi-Civita connection to the spin bundle which leads to a unique choice for the spin connection given by
 \begin{equation}\label{CanonicalSpinorConnectionR}
(\omega^\bg)_{\mu}{}^{ab}\equiv \eta^{ac}\viu{b}{\nu}\lr{\pa{\mu}\vid{c}{\nu}+\vid{c}{\al}{C_{\mu\al}}^\nu}
\end{equation}
where $C_{\mu\al}{}^\nu$ is the Levi-Civita connection of $\bg$, $\viu{a}{\mu}$ are the tetrads defined by  $g_{\mu\nu}=\viu{a}{\mu}\viu{b}{\nu}\eta_{ab}$, $\vid{a}{\mu}$ are its inverses $\vid{a}{\mu}\viu{a}{\nu}=\delta^\mu{}_\nu$ and we also have that $\ga^\mu=\vid{a}{\mu}\ga^a$ where $\ga^a$ are the flat Dirac mgamma matrices. The tetrads can be understood as relating a coordinate frame to a field of frames in the tangent bundle in which the metric looks Minkowskian. The affine connection can still be lifted in non-Riemannian space-times in a similar manner as in Riemannian space-times, leading to the cannonical spin connection given by
\begin{equation}\label{CanonicalSpinorConnection}
(\omega^{\bf\Ga})_{\mu}{}^{ab}\equiv \eta^{ac}\viu{b}{\nu}\lr{\pa{\mu}\vid{c}{\nu}+\vid{c}{\al}{\Ga_{\mu\al}}^\nu}.
\end{equation} 
However as pointed out in \cite{Hurley}, such lift is not sensitive to some of the irreducible components of the non-metricity tensor, and one could in principle choose a non-canonical spin connection by adding terms related to non-metricity and/or torsion by hand (see for instance \cite{Hurley,Janssen:2018exh}). Since we are concerned with minimal coupling, we will consider only the canonical piece of the spin connection, since any extra terms would change the form of $(\Upsilon_{NR})^\psi_\mu$ potentially adding non-minimal interactions.\\
 
 Going back to the minimal coupling discussion, we can see that in Riemannian space-times (where $\na=\na^\bg$), both actions \eqref{SpinorActionH} and \eqref{SpinorActionWrong} are equivalent since they only differ in a boundary term which is oblivious for the field equations:
\begin{equation}\label{RiemannianActionsBTerm}
\tilde{S}^R_\psi=S^R_\psi+\int_{\mathcal{V}}\dvg Div_g({\bf J}_\psi/2),
\end{equation}
with $ J_\psi^\mu=i\bpsi\ga^\mu\psi$ and where the superindex $R$ stands for the Riemannian. Notice that ${\bf J}_\psi$ is the Noether current \eqref{Noethercurrent} corresponding to a global $U(1)$ (or other Lie group) symmetry of the spinor action \eqref{SpinorActionH}. While $\tilde S_\psi^R$ is the traditional form of the action for spin 1/2 fields, the reason why $S_\psi^R$ is sometimes employed in Riemannian space-times is because it is explicitly hermitian (see e.g. \cite{BirrellDavies,ParkerBook}). However, in presence of torsion and/or non-metricity the two actions \eqref{SpinorActionH} and \eqref{SpinorActionWrong} for spin $1/2$ are not dynamically equivalent anymore, since they are no longer related by a boundary term, satisfying instead the relation 
\begin{equation}
\tilde{S}_\psi=S_\psi+\displaystyle\int_{\mathcal{V}}\dvg Div_g({\bf J}_\psi/2)-\displaystyle\int_{\mathcal{V}}\dvg\lr{\sigma_\mu J_\psi^\mu}\;,
\end{equation}
where again we have defined a non-Riemannian current
\begin{equation}\label{NRcurrent2}
\sigma_\mu\equiv{{Q_{[\al\mu]}}^\al+{S_{\al\mu}}}^\al=\Sigma_\mu-\frac{1}{2}Q_{\al\mu}{}^\al,
\end{equation}
which differs from the non-Riemannian current $\Sigma^\mu$ that appears in the scalar field case \eqref{NRcurrent} in a trace of the non-metricity tensor. Since both actions are not equivalent, we must take care of using \eqref{SpinorActionH} in non-Riemannian space-times, since it is the hermitic while \eqref{SpinorActionWrong} is not. However, we will see that there are further consistency reasons to choose \eqref{SpinorActionH} over \eqref{SpinorActionWrong}. To this end let us derive the field equations for $\psi$ and $\bpsi$ that each of both actions describe. Beginning with \eqref{SpinorActionH} and making use of \eqref{eom} we arrive to the field equations
\begin{align}\label{preeomspinor}
\begin{split}
&\lrsq{i\ga^\mu\na_\mu+\frac{i}{2}\lr{(\na_\mu\ga^\mu)+\ga^\mu\sigma_\mu-\frac{1}{2}{Q_{\al\mu}}^\al\ga^\mu}-m}\psi=0 ,\\
&\bpsi\lrsq{i\overleftarrow{\na}_\mu\ga^\mu+\frac{i}{2}\lr{\lr{\na_\mu\ga^\mu}+\ga^\mu\sigma_\mu-\frac{1}{2}{Q_{\al\mu}}^\al\ga^\mu}+m}=0,
\end{split}
\end{align}
where, since the gamma matrices have two spinorial indices and a Lorentz one, their covariant derivative is given by $\na_\mu\ga^\al=\pa{\mu}\ga^\al+{\Ga_{\mu\nu}}^\al\ga^\nu+[\ga^\al,\Upsilon^\psi_\mu]$ (see e.g \cite{Parker}). Using \eqref{CanonicalSpinorConnection}, we can compute $\Upsilon^\psi_\mu$ and then, by using the algebraic properties of the Dirac gamma matrices\footnote{We are using the conventions ${\ga^\mu}^\dagger=\dm^0 \ga^\mu\dm^0$, with ${\dm^0}^\dagger=\dm^0$ and ${\dm^0}^2=\mathbb{I}$. Notice the subtlety that this $\dm^0$ is not $\ga^\mu$ with $\mu=0$, but it has the same matrix form (see e.g. \cite{ParkerBook})}, we can compute the covariant derivative of the gamma matrices, finding
\begin{equation}\label{covdergamma}
 \na_\mu\ga^\al=\frac{1}{2}{Q_{\mu\nu}}^\al \ga^\nu.
\end{equation}
Plugging this result into the spinor field equations \eqref{preeomspinor} we have have that the action \eqref{SpinorActionH} leads to the following equations
\begin{align}\label{eomspinor}
\begin{split}
&\lrsq{i\ga^\mu\na_\mu+\frac{i}{2}\ga^\mu\sigma_\mu-m}\psi=0 ,\\
&\bpsi\lrsq{i\overleftarrow{\na}_\mu\ga^\mu+\frac{i}{2}\ga^\mu\sigma_\mu+m}=0.
\end{split}
\end{align}
Let us point out that these field equations were already found in \cite{Formiga} from the standard Riemannian equations in an alternative manner. Repeating the same procedure now with the non-hermitian action \eqref{SpinorActionWrong}, we are led to the field equations
\begin{align}\label{eomspinorWrong}
\begin{split}
&\lrsq{i\ga^\mu\na_\mu-m}\psi=0 ,\\
&\bpsi\lrsq{i\overleftarrow{\na}_\mu\ga^\mu+i\sigma_\mu\ga^\mu+m}=0.
\end{split}
\end{align}\\
A quick consistency check of both sets of field equations, \eqref{eomspinor} and\eqref{eomspinorWrong}, can be made by investigating if the on-shell relation between $\bpsi$ and $\psi$ is compatible with the group theoretical definition of adjoint spinor $\bpsi=\psi^\dagger\dm^0$. For that purpose, let us call $(\bpsi, \psi)$ and $(\bpsi,\psi)$ to a pair of arbitrary solutions of \eqref{eomspinor}  and \eqref{eomspinorWrong} respectively. Consider taking the adjoint of the equation satisfied by $\psi$:
\begin{equation}
\psi^\dagger\lrsq{-i(\overleftarrow{\pa{\mu}}-{\Upsilon^\psi_\mu}^\dagger){\ga^\mu}^\dagger-\frac{i}{2}\sigma_\mu{\ga^\mu}^\dagger-m}=0.
\end{equation}
Using the standard properties of the Dirac matrices, and the form of $\Ga_\mu$ given by  \eqref{CanonicalSpinorConnection}, we can show that ${\Upsilon^\psi_\mu}^\dagger=-\dm^0\Upsilon^\psi_\mu\dm^0$, which leads to
\begin{equation}\label{eomadjointconsistent}
\lr{\psi^\dagger\dm^0}\lrsq{i\overleftarrow{\na}_{\mu}\ga^\mu+\frac{i}{2}\sigma_\mu{\ga^\mu}+m}=0;
\end{equation}
where $\lr{\psi^\dagger\dm^0}\overleftarrow{\na}_\mu\equiv\lr{\psi^\dagger\dm^0}\lr{\overleftarrow{\pa{\mu}}+\Upsilon^\psi_\mu}$. Notice that the above equation is identical to the equation satisfied by $\bpsi$, thus consistently suggesting the identification $\bpsi=\psi^\dagger\dm^0$. Repeating the same procedure for the field equations defined by the action \eqref{SpinorActionWrong}, by taking the adjoint of the field equation satisfied by $\psi$ (i.e. the fist of \eqref{eomspinorWrong}) we arrive at
\begin{align}\label{eomadjointinconsistent}
\lr{\psi^\dagger\dm^0}\lrsq{i\overleftarrow{\na}_\mu{\ga^\mu}+m}=0;
\end{align}
which clearly differs from the equation satisfied by $\bpsi$ (i.e. the second of \eqref{eomspinorWrong}) due to the terms involving the non-Riemannian current $\sigma_\mu$. Therefore, the group-theoretical definition of $\bpsi=\psi\dm^0$ is not consistent with the dynamics given by the action \eqref{SpinorActionWrong} unles non-metricity and torsion vanish, when both \eqref{SpinorActionH} and \eqref{SpinorActionWrong} become dynamically equivalent. These findings imply that, in non-Riemannian space-times, the MCPL  has a consistent implementation only when applied to the Minkowskian version of the action \eqref{SpinorActionH} and not that of \eqref{SpinorActionWrong}. Thus, if generic non-minimal couplings between the spinor fields and torsion and/or non-metricity are considered, subtleties of this kind may arise if one does not apply the minimal prescriptions properly and to the correct Minkowskian action.\\

Since we know that within the MCPL only the action \eqref{SpinorActionH} is valid, let us now investigate whether spin $1/2$ fields couple explicitly to non-metricity and/or torsion within this action. As already pointed out in section \ref{sec:divopEL} for an arbitrary spin field, the only couplings between the spin 1/2 field and the non-Riemannian terms that arise in a minimal coupling prescription come from the form of the spinor connection. Here we will again proceed by using  \eqref{DecompConnSymb} to split the $\na_\mu\psi$ term in \eqref{eomspinor} into its Riemannian and non-Riemannian pieces. Assuming the canonical spin connection  \eqref{CanonicalSpinorConnection} we can use decomposition of the spacetime connection \eqref{DecompConnSymb} to find 
\begin{equation}\label{ContributionsOfSpinorConnection}
\begin{split}
&\ga^\mu\na_\mu\psi=\lrsq{\ga^\mu\na_\mu^\bg-T_\psi-\frac{1}{2}\sigma_\mu\ga^\mu}\psi\\
&(\na_\mu\bpsi)\ga^\mu=\bpsi\lrsq{\overleftarrow{\na}_\mu^\bg\ga^\mu+T_\psi-\frac{1}{2}\sigma_\mu\ga^\mu}\\
\end{split}
\end{equation}
where the Riemannian covariant derivatives act on spinor fields as $\na^\bg_\mu\psi=\lr{\ga^\mu\pa{\mu}-(\Upsilon^\bg)_\mu^\psi}\psi$ and on adjoint spinor fields as  $\bpsi\overleftarrow{\na}^\bg_\mu=\bpsi\lr{\overleftarrow{\pa{\mu}}+(\Upsilon^\bg)_\mu^\psi}$; and where we have defined 
\begin{align}
&(\Upsilon^\bg)_\mu^\psi=\frac{1}{2}{(\omega^\bg)_{\mu}}^{ab}\sigma_{ab}\\
&T_\psi=-\frac{i}{8}\epsilon^{abcd}S_{abc}\ga_d\ga_5.
\end{align}
Notice that $T_\psi$ is the well known interaction between the spinor fields and the totally-antisymmetric part of the torsion tensor \cite{Hehl6}. Using \eqref{ContributionsOfSpinorConnection}, the field equations for $\psi$ and $\bpsi$ derived from the right spinor action \eqref{SpinorActionH}, which are given by \eqref{eomspinor}  become
\begin{align}\label{eomspinordecomposed}
\begin{split}
&\lrsq{i\ga^\mu\na^{\bg}_\mu-i T_\psi-m}\psi=0 ,\\
&\bpsi\lrsq{i\overleftarrow{\na}^{\bg}_\mu\ga^\mu+i T_\psi+m}=0 .
\end{split}
\end{align}
Notice that only the Levi-Civita part of $\omega_{\mu}{}^{ab}$ and the totally antisymmetric part of the torsion appear in \eqref{eomspinordecomposed}. The non-metricity tensor and the other Lorentz-irreducible pieces of the torsion tensor do not couple (explicitly) to spin 1/2 field if the canonical spin connection \eqref{CanonicalSpinorConnection} is assumed. As in the case of the scalar field, a careful analysis of the action \eqref{SpinorActionH} (again decomposing the spinnor connection in its Levi-Civitta, torsion and non-metric parts) shows that the action only contains the Levi-Civita part of ${\omega_{\mu}}^{ab}$ and the totally antisymmetrized torsion tensor. Hence, no coupling to other pieces of the torsion tensor or to the non-metricity tensor should appear in the field equations either.\\

Let  us now analize the aplicability of our version of the MCPF to spin 1/2 fields. The Minkowskian field equations for a free spin 1/2 field and its adjoint written in a frame independent way are
\begin{align}\label{MCPFspinoreom}
\begin{split}
&\lrsq{i\ga^\mu\na^M_\mu-m}\psi=0\; ,\\
&\bpsi\lrsq{i\overleftarrow{\na}^M_\mu\ga^\mu+m}=0,
\end{split}
\end{align} 
where $\na^M$ stands for the covariant derivative of Minkowski space-time, {\it i.e.} the one associated to a purely inertial connection (which vanishes in any inertial reference frame). Following the MCPF, the non-Riemannian version of the equations is
\begin{align}\label{MCPFspinoreom}
\begin{split}
&\lrsq{i\ga^\mu\na_\mu-m}\psi=0\; ,\\
&\bpsi\lrsq{i\overleftarrow{\na}_\mu\ga^\mu+m}=0.
\end{split}
\end{align}
As we already computed, the adjoint of the first equation in \eqref{MCPFspinoreom} leads to the second equation of \eqref{MCPFspinoreom} if the identification $\bpsi=\bpsi^\dagger\ga^0$ is made, which is perfectly consistent with the group-theoretical definition of $\bpsi$. Therefore the MCPF procedure is, in principle, also consistent to describe a field belonging to the spin $1/2$ representation of the Lorentz group. Let us thus investigate the different couplings between spin 1/2 fields and the geometry if the MCPF, instead of the MCPL, is applied.  We can use  \eqref{ContributionsOfSpinorConnection} to re-write \eqref{MCPFspinoreom} as
\begin{align}\label{eomspinordecomposedMCPF}
\begin{split}
&\lrsq{i\ga^\mu\na^{\bg}_\mu-i T_\psi-\frac{i}{2}\sigma_\mu\ga^\mu-m}\psi=0 ,\\
&\bpsi\lrsq{i\overleftarrow{\na}^{\bg}_\mu\ga^\mu+i T_\psi-\frac{i}{2}\sigma_\mu\ga^\mu+m}=0,
\end{split}
\end{align}
which shows how within the MCPF, the non-Riemannian current $\sigma^\mu$ defined in \eqref{NRcurrent2} introduces a direct coupling between spin $1/2$ fields, torsion and non-metricity, therefore failing in being a minimal coupling prescription, as was the case for scalar fields. Experimentally probing this kind of couplings could help in elucidating wether the MCPF or the MCPL is more suited to describe spin 1/2 fields in non-Riemannian space-times. Nevertheless, following the same reasoning as for complex scalar fields, if we assume MCPF dynamics for spinor fields, which is given by \eqref{eomspinordecomposedMCPF}, a direct calculation shows that conservation of the fermionic current $J_\psi^\mu=i\bpsi\ga^\mu\psi$ is spoiled by the non-Riemannian current $\sigma^\mu$, having
\begin{equation}
Div_g({\bf J}_\psi)\big|_{MCPF}=\sigma_\mu  J_\psi^\mu,
\end{equation}
which could lead to charge violation through the non-Riemannian current $\sigma^\mu$. In contrast the MCPL action for spinor fields \eqref{SpinorActionH} is $U(1)$ invariant, which ensures conservation of ${\bf J}_\psi$ on-shell. Given that the global Standard Model symmetries is paramount in our current understanding of the universe, we can conclude that as for scalar fields, the MCPF is again pushed to an uncomfortable corner for describing spin $1/2$ field dynamics in presence of torsion and/or non-metricity. We believe that this solves the question raised in \cite{Formiga,Chen:2019zhj} (or at least shows a physical way to distinguish between both prescriptions) regarding whether the MCPF or MCPL is more appropriate for describing fermions in non-Riemannian spacetimes, favoring the MCPL if charge conservation is to be satisfied. Thus, assuming that charge conservation holds, the correct description for minimal coupling prescription for spin 1/2 fields in non-Riemannian space-times is the MCPL, which leads to the action 

\begin{equation}
S_\psi=\displaystyle\int_{\mathcal{V}}\dif V_\bg\lrsq{\frac{i}{2}\lr{\bpsi\ga^\mu(\na_\mu\psi)-(\na_\mu\bpsi)\ga^\mu\psi}+\bpsi m\psi}.
\end{equation}
Notice that, contrary to the findings of \cite{Chen:2019zhj}, the above action (which is their equation I.4 if we assume vanishing non-metricity) actually gives a consistent minimally coupling prescription for spin 1/2 fields, since it leads to the correct minimal coupling field equations
\begin{align}
\begin{split}
&\lrsq{i\ga^\mu\na^{\bg}_\mu-i T_\psi-m}\psi=0 ,\\
&\bpsi\lrsq{i\overleftarrow{\na}^{\bg}_\mu\ga^\mu+i T_\psi+m}=0,
\end{split}
\end{align}
and the covariant derivative needs no modification to achieve a consistent minimal coupling for spin 1/2 fields. Let us finally comment on the different role played by torsion in teleparallel theories and other theories such as the ECKS theory. In the later the torsion tensor is related the spin of the matter sources and it does not propagate new degrees of freedom, its sole role being to source a four-fermion contact interaction \cite{Kibble}. However, in the teleparallel framework curvature and non-metricity are set to zero and torsion is the only non-vanishing geometrical object which encodes relevant information about the propagation of gravitational degrees of freedom. Thus, while the torsion coupling to fermions given by $\Upsilon_S^\psi$  only encodes contact interactions in the context of theories like the ECKS, in the teleparallel framework it could contain an additional coupling to the gravitational field which might be worth exploring in future works.

\subsection{The minimally coupled vector field}\label{sec:vecfield}

Let us now finish by study whether the different minimal coupling prescriptions for massless vector fields make sense\footnote{The following argumentation applies trivially to massive vector fields.}. Since the Minkowskian action for massless vector fields is constructed only with the gauge invariant 2-form $\boldsymbol{F}=\dif \boldsymbol{A}$, which is a frame independent object, following the MCPL as stated in section \ref{sec:MinCoup}, the action for a minimally coupled gauge field in an arbitrary non-Riemannian space-time is given by
\begin{equation}\label{vectoraction}
S_A=\displaystyle -\frac{1}{4} \int_\mathcal{V}\dvg g^{\mu\al}g^{\nu\be} F_{\mu\nu}F_{\al\be}\;,
\end{equation}
with $F_{\mu\nu}=(\dif\boldsymbol{A})_{\mu\nu}$. In order to apply the EL equations \eqref{eom} to the above action, we need to re-write it as a functional of $\{\boldsymbol{A},\na \boldsymbol{A}\}$. We can do this by using the relation between the exterior derivative and the covariant derivative, thus finding
\begin{equation}\label{vectoractionnice}
\begin{split}
S_A=\displaystyle - \frac{1}{4}\int_\mathcal{V}&\dvg g^{\mu\al}g^{\nu\be} \big[4\na_{[\mu}A_{\nu]}\na_{[\al}A_{\be]}-\\&-4(\na_{[\mu}A_{\nu]})S_{\al\be}{}^\sigma A_\sigma+S_{\mu\nu}{}^\rho S_{\al\be}{}^\sigma A_\rho A_\sigma \;.
\end{split}
\end{equation}
 With the action written in this form, we can directly apply \eqref{eom} and obtain the field equations for the massless vector field in non-Riemannian spaces following the MCPL, which are
\begin{equation}\label{preeomvector}
\na_\mu F^{\mu\nu}-\lr{\frac{1}{2}{Q_{\mu\al}}^\al+{S_{\mu\al}}^\al}F^{\mu\nu}+\frac{1}{2}{S_{\al\be}}^\nu F^{\al\be}=0.
\end{equation}
 Again, by using \eqref{DecompConnSymb}, we can decompose the covariant derivative in its Riemannian and non-Riemannian parts, finding the identity
\begin{equation}\label{DecompCovDerFmunu}
\na_\mu F^{\mu\nu}=\na^\bg_\mu F^{\mu\nu}+\lr{\frac{1}{2}{Q_{\mu\al}}^\al+{S_{\mu\al}}^\al}F^{\mu\nu}-\frac{1}{2}{S_{\al\be}}^\nu F^{\al\be},
\end{equation} 
which plugged back on the field equations \eqref{preeomvector} leads to the well known
\begin{equation}\label{eomvector}
\na^{\bg}_\mu F^{\mu\nu}=0\;.
\end{equation}
which can also be written as
\begin{equation}\label{eomvector}
\bdel_g\boldsymbol{F}=0\;.
\end{equation}
or in terms of the vector field
\begin{equation}\label{EqvectorMCPL}
\bdel_g\dif\boldsymbol{A}=0.
\end{equation}
Since neither $\dif$ nor $\bdel_g$ know about the affine structure, the result is that as happens for scalar fields, vector fields that are minimally coupled according to our version of the MCPL do not couple explicitly either to torsion and/or to non-metricity explicitly\footnote{Minimally coupled vector fields can also feel non-metricity-related effects through the metric, see \cite{Letter,Delhom:2019wir})}. Let us now see what happens if we take the naive prescription $\partial\mapsto\nabla$ too seriously. This leads by the naive MCPL to the non-Riemannian Lagrangian
\begin{equation}\label{vectoractionWrong}
S_A=\displaystyle - \int_\mathcal{V}\dvg g^{\mu\al}g^{\nu\be}\na_{[\mu}A_{\nu]}\na_{[\al}A_{\be]}\;.
\end{equation}
Note that whereas the action \eqref{vectoraction} obtained by applying our vesrion of the MCPL is invariant under projective transformations $\Gamma_{\mu\nu}{}^\al\to\Gamma_{\mu\nu}{}^\al+\xi_\mu \delta^\al{}_\nu$, while the above action resulting of applying the naive MCPL is not. Projective symmetry has been recently proven to be relevant in order to avoid ghost degrees of freedom in the construction of metric-affine theories of gravity \cite{GhostsLargo,BeltranJimenez:2019acz,Aoki:2019rvi}. { Also, as it is well known, this non-minimal coupling also breaks the gauge invariance of the spin-1 kinetic term in a non-trivial torsion background.} Last but not least, this coupling could also introduce potential violations of the equivalence principle, given that the trajectories of the photons in the eikonal limit would suffer deviations from the geodesic ones due to a torsion-induced fifth force. {Given that no violations of gauge invariance or deviations from geodesic trajectories have been detected so far, this prescription is pushed to an uncomfortable corner when compared to experimental data \cite{Ni:2009fg,Ni:2015poa,AudretschLammerzahl_1983,Lammerzahl:1997wk,Mohanty:1998vs}. Furthermore, since it introduces a coupling between the vector fields and the torsion which is not present in our proposal for minimal coupling, we argue that this is not a minimal-coupling prescription according to our definition (see sec. \ref{sec:MinCoup}). Let us nonetheless derive the field equations corresponding to the above action \eqref{vectoractionWrong} in order to see explicitly the appearances of these couplings that do not appear when our minimal coupling prescription is employed. }By using again \eqref{eom} in the naive MCPL action, we arrive to the field equations
\begin{equation}\label{preeomvector}
\na_\mu F^{\mu\nu}-\lr{\frac{1}{2}{Q_{\mu\al}}^\al+{S_{\mu\al}}^\al}F^{\mu\nu}=0,
\end{equation}
which after splitting the covariant derivative can be recast into
\begin{equation}\label{eomvector}
\na^{\bg}_\mu F^{\mu\nu}-\frac{1}{2}S_{\mu\al}{}^\nu F^{\mu\al}=0\;,
\end{equation}
or also
\begin{equation}\label{eomvector}
\bdel_g\boldsymbol{F}-\frac{1}{2}S_{\mu\al}{}^\nu F^{\mu\al}=0
\end{equation}
which can also be written
\begin{equation}
\bdel_g\dif\boldsymbol{A}-\frac{1}{2}S_{\mu\al}{}^\nu (\dif\boldsymbol{A})_{\mu\al}=0. 
\end{equation}
Again, since neither $\dif$ nor $\bdel_g$ know about the affine structure, there can be no cancellation of the torsion terms, and therefore the naive MCPL leads to a non-minimal coupling between $\boldsymbol{A}$ and the torsion tensor, which makes the naive MCPL fail as a minimal coupling prescription.\\

Let us now investigate the MCPF and naive MCPF prescriptions. When written in a frame independent way, the source free vector field equations in Minkowski space-time are 
\begin{equation}\label{preeomvectorMCPF}
\boldsymbol{\delta}_{{\eta}} \boldsymbol{F}=0\quad,\quad \boldsymbol{F}=\dif\boldsymbol{A}
\end{equation}

where $\boldsymbol{\delta}_{{\eta}}$ is the codiferential operator associated to the Minkowski metric $\boldsymbol{\eta}$ (see footnote \ref{footnotedalambertian}), and in coordinates it reads $(\boldsymbol{\delta}_{{\eta}} \boldsymbol{F})^\mu=\partial_\al F^{\al\mu}$ and $F_{\mu\nu}=\partial_{[\mu}A_{\nu]}$. Applying the MCPF as defined in section \ref{sec:MinCoup}, in a general space-time we must use the codiferential operator associated to the space-time metric $\bg$, so that acording to the MCPF the vector field equations in a general space-time read
\begin{equation}\label{preeomvectorMCPF}
\boldsymbol{\delta}_g \boldsymbol{F}=0\quad,\quad \boldsymbol{F}=\dif\boldsymbol{A}
\end{equation}
and in coordinates it is satisfied $(\boldsymbol{\delta}_g \boldsymbol{F})^\nu=\nabla_\mu^\bg F^{\mu\nu}$, thus having again that the MCPF is consistent with the MCPL even in presence of non-metricity and/or torsion. In the Lorenz gauge, which in Minkowski space-time is characterised by $\boldsymbol{\delta}_{{\eta}}\boldsymbol{A}=0$  and in general space-times is characterised by In the Lorenz gauge, characterised by $\boldsymbol{\delta}_{{g}}\boldsymbol{A}=0$, the field equations are  
\begin{equation}
\Box_{{\eta}} \boldsymbol{A} = 0\qquad\text{and}\qquad \Box_g \boldsymbol{A} = 0,
\end{equation}
respectively. Since we are in the Lorentz gauge, then  $\Box_g\boldsymbol{A}=(\dif\bdel_g+\bdel_g\dif)\boldsymbol{A}=\bdel_g\dif\boldsymbol{A}$, thus recovering the same field equations as with our MCPL \eqref{EqvectorMCPL}. The last equation can also be written in the familiar form $\na^\bg_\mu \na^\bg{}^\mu A^\al+R_\bg{}^\al{}_\mu A^\mu=0$, where $R_\bg{}^\al{}_\mu$ is the Ricci tensor associated to $\bg$. Let us now see what would have ended up with had we applied the naive MCPF. Given that the Minkowskian field equations in any coordinates read
\begin{equation}\label{eomvectorMink}
\pa{\mu}{F}^{\mu\nu}=0\qquad , \qquad {F}_{\mu\nu}=2\pa{[\mu}A_{\nu]}
\end{equation}
by the naive prescription $\partial\to\na$, the naive MCPF would have lead us to the non-Riemannian field equations
\begin{equation}\label{eomvectorMCPF}
\na_\mu \tilde{F}^{\mu\nu}=0\qquad , \qquad \tilde{F}_{\mu\nu}=2\na_{[\mu}A_{\nu]}
\end{equation}
which after using \eqref{DecompCovDerFmunu} and the identity $\tilde{F}_{\mu\nu}=F_{\mu\nu}+S_{\mu\nu}{}^\al A_\al$ can be written as
\begin{align}
&\na^{\bg}_\mu F^{\mu\nu}+\na^\bg_\mu(S^{\mu\nu\al}A_\al)-\frac{1}{2}{S_{\al\be}}^\nu (F^{\al\be}+S^{\al\be\ga}A_\ga)+\nonumber\\
&+\lr{\frac{1}{2}{Q_{\mu\al}}^\al+{S_{\mu\al}}^\al}(F^{\mu\nu}+S^{\mu\nu\al}A_\al)=0.
\end{align}
or also
\begin{align}
&2\na^{\bg}_\mu \na^\bg{}^{[\mu}A^{\nu]}+\na^\bg_\mu(S^{\mu\nu\al}A_\al)-\frac{1}{2}{S_{\al\be}}^\nu (2\na^\bg{}^{[\al}A^{\be]}+S^{\al\be\ga}A_\ga)+\nonumber\\&+\lr{\frac{1}{2}{Q_{\mu\al}}^\al+{S_{\mu\al}}^\al}(2\na^\bg{}^{[\mu}A^{\nu]}+S^{\mu\nu\al}A_\al)=0.
\end{align}
In the Lorentz gauge, which by the naive MCPF would be characterized by $\na_\mu A^\mu$ this looks as
\begin{align}
&\Box_g A^{\nu}-\na^\bg{}^\nu\lrsq{\lr{S_{\al\mu}{}^\mu+\frac{1}{2}Q_{\al\mu}{}^\mu}A^\al}+\na^\bg_\mu(S^{\mu\nu\al}A_\al)+\nonumber\\&+\lr{\frac{1}{2}{Q_{\mu\al}}^\al+{S_{\mu\al}}^\al}(2\na^\bg{}^{[\mu}A^{\nu]}+S^{\mu\nu\al}A_\al)-\\&-\frac{1}{2}{S_{\al\be}}^\nu (2\na^\bg{}^{[\al}A^{\be]}+S^{\al\be\ga}A_\ga)=0.\nonumber
\end{align}
We can see that this equations features several non-minimal coupling terms between torsion and non-metricity, thus showing one more time how the naive MCPF is not a minimal coupling prescription in the sense defined in section \ref{sec:MinCoup}.

\section{Outlook}\label{sec:Final}
We have here dealt with the issue of minimal coupling between matter fields and the geometry in presence of torsion and/or non-metricity. Since this is a confusing issue in the literature, we first gave a definition of what we understand as a minimal coupling prescription and what should not be regarded as such. Thus, in a heuristic sense, minimal coupling prescriptions are those that change as little as possible the differentiable operators appearing in the matter field equations and/or actions. Therefore, to implement a minimal coupling prescription consistently, one should pay attention to what differential operator that one is using in the Minkowskian theory, which is never $\partial$ since it is a different operator in non-inertially related frames (see section \ref{sec:MinCoup}), but rather $\dif$, $\bdel$ or $\na$; and be consistent when going to a general space-time in using the same operator. We note that this prescription reduces to the usual one in Riemannian space-times \cite{Will_Book}, but it is different from the naive prescription $\partial\to\nabla$ in non-Riemannian space-times in that it does not introduce the additional non-minimal interactions between matter and geometry that arise when the naive recipe is employed. Indeed, as we showed, if one implements the minimal coupling prescriptions in this way, the MCPF and MCPL are equivalent for spin 0 and 1 fields , while only the MCPL gives a consistent minimal coupling prescription for spinor fields. This results are in contradiction with \cite{Chen:2019zhj} due to the fact that the naive $\partial\to\na$ substitution is applied there to implement MCPF and MCPL in their respective naive versions. In the spin 1/2 case we also showed that charge is conserved only for the MCPL prescription, thus giving a possible way out to the dilemma found in \cite{Formiga} about whether one should use MCPF or MCPL for spin 1/2 fields in non-Riemannian space-times. As a by-product, we also showed how the non-hermitian action $\eqref{SpinorActionWrong}$ that is commonly used in Riemannian space-times to describe spinor fields is not equivalent to the hermitian one in presence of torsion and/or non-metricity, while in Riemannian space-times both give the same dynamics. As anticipated in section \ref{sec:MinCoup}, we also showed that the naive MCPL and naive MCPF typically fail in being minimal coupling prescriptions for all matter fields, the exception being the naive MCPL for the scalar field. This exception is due to the fact that only first order derivatives of the scalar enter in the action, and for scalar field the operators $\dif$ and $\na$ are by definition the same operator. We hope that this discussion can thus be useful regarding the issue of coupling matter fields to torsion and/or non-metricity (see {\it e.g.} \cite{Harko:2018gxr,Lobo:2019xwp,Xu:2019sbp}). As a final remark, let us add on how to understand this under the lens of the two frames that arise in several metric-affine modifications of GR. Namely, in some curvature-based modifications to GR, there are the Jordan and Einstein frames in which one can formulate the theory and if the matter is {\it minimally coupled} in one, it will not be in the other. However, one should note that the meaning of minimal coupling here is not the same one that we are discussing. While we have discuss minimal coupling between matter fields and the affine connection, the {\it non-minimal} couplings that appear on passing from one frame to the other do not involve the affine connection, and they only introduce new interactions between the fields of the matter sector.

\begin{acknowledgements}
A.D. is supported by  an FPU fellowship. This work is supported by the the Spanish Projects No. FIS2017-84440-C2-1-P (MINECO/FEDER, EU), the Project No. H2020-MSCA-RISE-2017 Grant No. FunFiCO-777740, Project No. SEJI/2017/042 (Generalitat Valenciana), the Consolider Program CPANPHY-1205388, and the Severo Ochoa Grant No. SEV-2014-0398 (Spain). I also thank Joan  Josep  Ferrando,  Juan  Antonio  Morales-Lladosa, Jose Beltr\'an Jim\'enez, Gonzalo J. Olmo, and Alejandro Jim\'enez Cano for useful discussions and comments during the elaboration of this work.
\end{acknowledgements}

\begin{widetext}
\section*{Appendix: Summary of the notation}

\begin{center}
    \begin{tabular}{ | l | l |}
    \hline
    {\bf Symbol} & {{\bf Meaning}} \\ \hline
    $\mathcal{M}$ & n-dimensional space-time manifold. \\ \hline 
    $\lrcorner$ & Interior product on $\mathcal{M}$. \\ \hline
    $\mathfrak{X}$ & Set of vector fields on $\mathcal{M}$. \\ \hline
    $\bg$ & A metric strucure of $\mathcal{M}$. \\ \hline 
    $g$ & Determinant of the metric $\bg$. \\ \hline 
    $\epsilon$ & Generic volume form on $\mathcal{M}$. \\ \hline
    $\dvg$ & Volume form associated to the metric $\bg$. In a coordinate frame it reads $\dvg=\sqrt{-g} dx^{\mu_1}\wedge...\wedge dx^{\mu_n}$. \\ \hline 
    $\star_\epsilon$ & Hodge dual operator associated to the volume form $\varepsilon$. It acts on differential forms on $\mathcal{M}$.\\ \hline
    {\bf d} & Exterior derivative of differential forms on $\mathcal{M}$. \\ \hline
    $\delta_\epsilon$ & Co-differential operator associated to the volume form $\varepsilon$, defined by $\delta_\epsilon\equiv \star_\epsilon\text{\bf d} \star_\epsilon$. \\ \hline
    $\Box_\varepsilon$ & D'Alambertian operator associated to the volume form $\epsilon$, defined by $\Box_\epsilon\equiv \textbf{d}\delta_\epsilon+\delta_\epsilon\textbf{d}$. \\ \hline
    $Div_\epsilon$ & Divergence operator on $\mathcal{M}$. It can be defined as in \eqref{Divop} or as $Div_\epsilon=\star_\epsilon \text{\bf d} \star_\epsilon$.\\ \hline
    $\partial$ & Formal symbol meaning a general partial derivative of a tensor or spinor field without the need for specifying a frame. \\ \hline
    $\partial_\mu$ & Partial derivative associated to the coordinate frame $\{x^\mu\}$  on $\mathcal{M}$. \\ \hline 
     ${\bf\Ga}$ & Affine connection.  \\ \hline
     $\nabla$ & Covariant derivative associated to an affine connection ${\bf \Ga}$.  \\ \hline
     $\Ga_{\mu\nu}{}^\al$ & Connection symbols associated to the affine connection ${\bf \Ga}$ (typically associated to the tensorial representations).\\ \hline
     $\nabla^M$ & Covariant derivative of Minkowski space-time. In a cartesian inertial frame it coincides with $\partial$.  \\ \hline
     $\nabla^\bg$ & Covariant derivative associated to the Levi-Civvita connection of \bg.\\ \hline
     $\textbf{C}(\bg)$ & Levi-Civvita connection of \bg. Its connection coefficients are given by $C_{\mu\nu}{}^{\alpha}$.  \\ \hline
        $\Psi_i$ & Collection of matter fields labelled by $i$, each of them belonging to an arbitrary representation of the Lorentz gorup \\ \hline
         $\Phi$ & Scalar field. \\ \hline
         $\psi$  & Spin $1/2$ field.\\ \hline
         $A^\mu$ & Spin $1$ field.\\ \hline
         $\Upsilon^i_\mu$ & Connection coefficients of ${\bf \Ga}$ in the representation corresponding to the matter field $\Psi_i$.  \\ \hline
         $(\Upsilon^i_{NR})_\mu$ & Non-Riemannian piece of $\Upsilon^i_\mu$.  \\ \hline
         $\Upsilon^\psi_\mu$ & Connection coefficients of ${\bf \Ga}$ in the spin $1/2$ representation.  \\ \hline
         $(\Upsilon^\bg)^\psi_\mu$ & Piece of $\Upsilon^\psi_\mu$ containing only the Levi-civitta part of ${\bf\Ga}$.  \\ \hline
         $(\omega^{\bf\Ga})_\mu{}^{ab}$ & Spin connection associated to the affine connection ${\bf\Ga}$.  \\ \hline
         $(\omega^{\bg})_\mu{}^{ab}$ & Spin connection associated to \textbf{C}(\bg), {\it i.e.} to the Levi-Civitta connection of \bg.  \\ \hline
         $T_\psi$ & Term encoding the interaction between the torsion tensor and a minimally coupled spin $1/2$ field.  \\ \hline

    \end{tabular}
\end{center}
\end{widetext}

\end{document}